\documentclass{aa}
\usepackage{xcolor}
\usepackage{natbib}
\usepackage{graphicx}
\usepackage{stfloats}
\usepackage{placeins}
\usepackage[flushleft]{threeparttable}
\usepackage{tikz}
\newcommand*\circled[1]{\tikz[baseline=(char.base)]
{\node[shape=circle,draw,inner sep=2pt] (char) {#1};}}
\usepackage{txfonts}
\usepackage[breaklinks,colorlinks,urlcolor=blue,citecolor=blue,linkcolor=blue]{hyperref}
\begin{document}

   \title{Simultaneous gas accretion onto a pair of giant planets: \\ Impact on their final mass and on the protoplanetary disk structure}

   \author{C. Bergez-Casalou\inst{1}, B. Bitsch\inst{1}, S. N. Raymond\inst{2}}

   \institute{\inst{1}Max-Planck-Institut für Astronomie, Königstuhl 17, 69117 Heidelberg, Germany, email:bergez@mpia.de\\
   \inst{2}Laboratoire d’Astrophysique de Bordeaux, CNRS and Université de Bordeaux, Allée Geoffroy St. Hilaire, 33165 Pessac, France \\}

%   \date{Received September 15, 1996; accepted March 16, 1997}

% \abstract{}{}{}{}{} 
% 5 {} token are mandatory
 
 \abstract
  {Several planetary systems are known to host multiple giant planets. However, when two giant planets are accreting from the same disk, it is unclear what effect the presence of the second planet has on the gas accretion process of both planets. In this paper we perform long-term  2D isothermal hydrodynamical simulations (over more than 0.5 Myrs) with the \texttt{FARGO-2D1D} code, considering two non-migrating planets accreting from the same gaseous disk. We find that the evolution of the planets' mass ratio depends on gap formation. However, in all cases, when the planets start accreting at the same time, they end up with very similar masses (0.9 $<m_{p,out}/m_{p,in}<$ 1.1 after 0.5 Myrs). Delaying the onset of accretion of one planet allows the planets' mass ratio to reach larger values initially, but they quickly converge to similar masses afterward (0.8 $<m_{p,out}/m_{p,in}<$ 2 in $10^5$ yrs). In order to reproduce the more diverse observed mass ratios of exoplanets, the planets must start accreting gas at different times, and their accretion must be stopped quickly after the beginning of runaway gas accretion (less than 0.5 Myrs), for example via disk dispersal. The evolution of the planets' mass ratio can have an important impact on the dynamics of the system and may constrain the formation history of Jupiter and Saturn.} 
  %leave it empty if necessary  


   \authorrunning{C.Bergez-Casalou et al}
   \titlerunning{Constraining planet formation during the disk phase}

   \maketitle
%
%________________________________________________________________

\section{Introduction}

Recent surveys show that planetary systems that host multiple giants are common \cite[e.g.,][]{Lissauer2012,Fabrycky2014,Zhu2022}. These planets are believed to have formed in the same protoplanetary disk, where they acquired their massive gaseous atmospheres. Previous hydrodynamical studies investigate either the growth of single planets in the disk \citep[e.g.,][]{Ayliffe2009,DAngelo2013,Crida2017,Schulik2019,Bergez2020} or the evolution of already formed multiple planets \citep[e.g.,][]{Baruteau2013,Lega2013,Pierens2014,Morbidelli2018}. This dichotomy originates from the difficulty to accurately model each evolution process.

It is vital to understand exactly how gas is accreted onto giant planets. Some studies have described how gas accretion can be approximated in 2D, with or without planet migration \citep{Kley1999,Crida2017}. Other studies used complex 3D simulations to take the various fluid and thermal processes governing the gas accretion of an embedded planet  into account \citep[e.g.,][]{Ayliffe2009,Machida2010,DAngelo2013,Szulagyi2017,Lambrechts2019,Schulik2019}. Due to their high computational cost, such simulations are often integrated over short timescales, around 100 planetary orbits, making it impossible to investigate the long-term growth of single accreting planets with the current computing facilities.

Moreover, the gas distribution around embedded giant planets is impacted by gap opening. The formation of these gaps has been observed both in the dust \citep[e.g.,][]{ALMA2015,Andrews2018,Benisty2021} and in the gas \citep[e.g.,][]{Huang2018b,Smirnov-Pinchukov2020}. Several studies have investigated the characteristics of the gap (such as its depth and width) as a function of the planet and disk characteristics \citep{LinPapaloizou1986,Crida2006,Fung2014,Kanagawa2015}. However, due to computational limitations, these studies neglected gas accretion. Recent works show that gap opening and gas accretion are highly dependent on each other through the viscosity of the disk \citep{Bergez2020,Rosenthal2020}. 

As gap opening has a non-negligible impact on the gas structure, the growth of a giant planet must be impacted by the presence of a second accreting planet in the same disk. The goal of this study is to quantify this impact as a function of the disk parameters and planet characteristics (i.e., their radial separation and delay in accretion). We perform 2D isothermal hydrodynamical simulations similar to \cite{Bergez2020} to monitor the growth of two planets accreting in the runaway accretion regime from the same disk. Our 2D isothermal setup allows us to integrate the evolution of the planet masses for around 0.5 Myrs, which is longer than the majority of the studies that investigate gas accretion onto planets using hydrodynamical simulations.

This paper is structured as follows: The numerical setup is described in Sect. \ref{sec:2pnumericalsetup}. In Sect. \ref{sec:2p_diffvisco} we investigate the impact of different disk viscosities on the growth of each planet. A comparison with single accreting planets is presented in Sect. \ref{sec:2p_2p1pcomp}. Different planet separations are investigated in Sect. \ref{sec:2p_diffseparation}. In Sect. \ref{sec:delayedaccretion} we show the mass evolution of planets accreting at different times. Our findings are discussed in Sect. \ref{sec:2p_discuss}, where we elaborate on the impact on the stellar gas accretion and on the dynamical evolution of the planetary systems. A comparison with the exoplanet population is discussed as well before we summarize and conclude in Sect. \ref{sec:2p_conclusion}.

\section{Numerical setup}
\label{sec:2pnumericalsetup}

In this paper we simulate two accreting planets on fixed circular orbits embedded in their gaseous disk with the hydrodynamical code \texttt{FARGO-2D1D} \citep{Crida2007}. This code is composed of two different kind of grids: a 2D grid where the planets are located and which is surrounded by two 1D grids. The first 1D inner disk ranges from 0.1 AU to 0.78 AU, the 2D grid ranging from 0.78 AU to 23.4 AU and the second outer 1D disk spans from 23.4 AU to 260 AU. With this code, the global viscous evolution of the disk is self-consistently modeled at a reasonable computational cost. As mentioned in \cite{Bergez2020}, it is important to have a consistent viscous evolution of the full disk in order to accurately describe gap formation and gas accretion.

For computational accuracy, the code uses dimensionless units. We normalized masses with the mass of the central star, $M_0 = M_\odot$ and lengths with the position of the planet, $r_0 = 5.2$ AU and we set the gravitational constant as $G=1$. The unit of time is therefore based on the orbital period at $r_0$ with $P = 2\pi t_0$ where $t_0 = (r_0^3/(GM_{*}))^{-1/2}$.

In order to focus on the impact of gas accretion, dynamical interactions between the planets and planet migration are neglected. This choice is discussed in Sect. \ref{sec:2p_planetdynamics}. The planets are fixed on circular orbits, at key positions corresponding to different period ratios. We use the term "period ratio" instead of "resonance," as they are not dynamically locked in resonance but are forced by the code to stay at their position. Four period ratios are investigated: 2:1, 3:1, 4:1, and 5:1. These positions were chosen such as the planets are far enough from each other to be considered dynamically stable during their growth \citep{Chambers1996,Raymond2009a}. In order to compare to a single-planet case, we also simulate the growth of single planets located at positions corresponding to the investigated period ratios. The planets' configurations are summarized in table \ref{tab:2p_planetconfig}.

Each planet starts with an initial mass of 20 $\rm M_\oplus$, allowing it to directly accrete in the runaway gas accretion regime \citep{Pollack1996}. These initial cores are slowly introduced in the disk with the following mass-taper function:
\begin{equation}
    m_{\rm taper} = \sin^2{(t/(4 n_{\rm orb}))}
    \label{masstapereq}
,\end{equation}
making the planet grow from 0 to its initial mass in $n_{orb} =3$ orbits of the inner planet. Except for Sect. \ref{sec:delayedaccretion} where the accreting times are specified, all planets start accreting simultaneously after $100$ orbits of the inner planet. 

The planets accrete following the accretion routine described in \cite{Bergez2020}. With this accretion routine, the amount of gas accreted by the planets is dictated by \cite{Machida2010} ($\dot{M}_{M}$) and is limited to what the disk can provide by the approach of \cite{Kley1999} ($\dot{M}_{K}$). It is written as follows:
\begin{equation}
    \dot{M}_p = min
    \begin{cases}
      \dot{M}_{M}=<\Sigma>_{0.9r_H} H^2\Omega_K \times min\Big[0.14;0.83(r_H/H)^{9/2}\Big]\\
      \dot{M}_{K}= \iint_{A_{disk}} f_{\rm red}(d)\;\Sigma(r,\phi,t)\;\pi f_{\rm acc} \; dr \; d\phi
    \end{cases}      
    \label{eq:accroutine}
,\end{equation}
where $<\Sigma>_{0.9r_H}$ is the averaged surface density around the planet up to 0.9 $r_H$ with $r_H = r_p(m_p/3M_*)^{1/3}$ being the Hill sphere of the planet; $A_{disk}$ is the disk area; $H$ is the disk scale height; $\Omega_K$ is the Keplerian orbital frequency of the planet; $d$ is the distance from the planet; $f_{\rm acc}$ is the inverse timescale upon which the accretion rate of \cite{Kley1999} is occurring; and $f_{\rm red}$ is a smooth reduction function that predicts what fraction of gas must be accreted as a function of the distance from the planet. Values for $f_{\rm acc}$ and $f_{\rm red}$ are chosen in order to reproduce a realistic accretion efficiency\footnote{Note that $<\Sigma>_{0.9r_H}$ is averaged until 0.9 $r_H$ because $f_{\rm red}=0$ for $d>0.9r_H$ in our case.}. The detailed accretion routine is presented in Appendix \ref{appendix:gas_accretion_routine}.

Once the amount of gas is determined by the accretion routine, it is removed from the disk and added to the planet's mass as in \cite{Kley1999}. Unless specified, we remain in the regime where $\rm \dot{M}_M < \dot{M}_K$ throughout, meaning that we always remove the amount of mass suggested in \cite{Machida2010}. 

The resolution of the 2D grid is such that there are five cells per Hill radius of the inner planet before it starts growing, which leads to $N_r = 802$ and $N_\phi = 1158$. Considering that the resolution is fixed in time, the Hill sphere region will become better and better resolved as the planets grow ($\rm r_{\rm H} \propto m_{\rm p}^{1/3}$). As we enhance the planet separation, we need to adjust the 2D-1D boundary located at the outer edge of the 2D grid in order to properly take into account the perturbations of the furthest planet. Therefore, in the 4:1 and 5:1 case, this boundary is moved to 36.4 AU, enhancing the radial number of cells of the 2D grid to $N_r = 1262$. The azimuthal number of cells remains unchanged. 

\begin{table}[t]            % title of Table
\centering                          % used for centering table
\caption{Semimajor axis of the different planet configurations considered in this paper.} 
\begin{tabular}{c c c c}        % centered columns (4 columns)
\hline             
Configuration & Inner planet $r_p$ & Outer planet $r_p$ \\   
\hline                  
    2 planets 2:1 &  5.2 AU & 8.22 AU \\
    2 planets 3:1 &  5.2 AU & 10.82 AU \\
    2 planets 4:1 &  5.2 AU & 13.42 AU \\
    2 planets 5:1 &  5.2 AU & 15.18 AU \\
\hline
    1 planet inner &  5.2 AU & - \\
    1 planet outer 2:1 &  - & 8.22 AU \\
    1 planet outer 3:1 &  - & 10.82 AU \\
    1 planet outer 4:1 &  - & 13.42 AU \\
    1 planet outer 5:1 &  - & 15.18 AU \\
\hline
    
\end{tabular}
\label{tab:2p_planetconfig}
\end{table}

In \texttt{FARGO-2D1D}, the disk is locally isothermal. As in \cite{Bergez2020}, the surface density profile is chosen such that the total mass of the disk is $M_d = 0.1 M_*$, leading to $\Sigma_0 (r_0) = 93.6 \; \mathrm{g/cm^2}$ with $r_0$ = 5.2 AU. Even if this can be considered as a heavy disk \citep{Baillie2019}, its large radial extent allows us to neglect self-gravity. The aspect ratio, $h=H/r$, of the disk is constant. The disk is subject to an $\alpha$-viscosity as described by \cite{Shakura1973}. In the following section, different gas kinematic viscosities $\nu = \alpha h^2 r^2 \Omega_K$ are investigated by varying the aspect ratio as well as the $\alpha$-viscosity parameter. The investigated values are $h=0.03,\mathbf{0.05},0.07$ and $\alpha=10^{-4},\mathbf{10^{-3}},10^{-2}$, with our fiducial values marked in bold.

\section{Influence of the disk viscosity}
\label{sec:2p_diffvisco}

The flow of gas in the disk is dictated by the kinematic viscosity, $\nu$, which depends on the $\alpha$-viscosity parameter on one hand and on the aspect ratio $h$ of the disk on the other. In this section we investigate the influence of the disk viscosity on the accretion behavior of two planets fixed in a 3:1 period ratio. We start by changing the $\alpha$-viscosity parameter in Sect. \ref{sec:2p_diffalphas}, then the influence of different aspect ratios is studied in Sect. \ref{sec:2p_diffh}. We focus on the influence of the viscosity on the evolution of the planet mass-ratio in a last subsection (Sect. \ref{sec:2p_massratiosdiffnu}).

\subsection{Influence of the turbulent viscosity}
\label{sec:2p_diffalphas}

\begin{figure*}[t]
        \centering   
        \includegraphics[width=18cm]{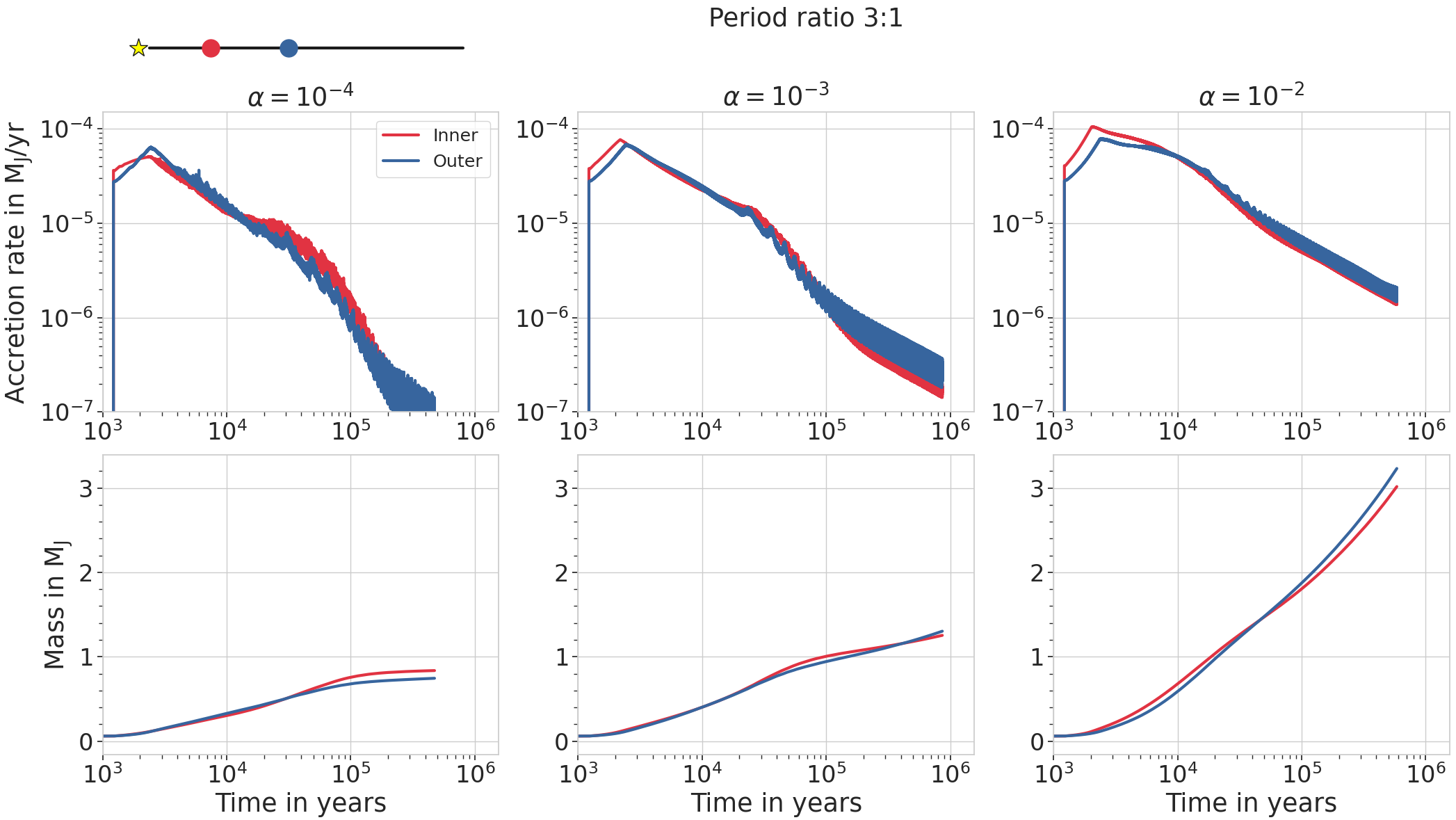}
        \caption[Planetary accretion rates and masses as a function of time for different $\alpha$-viscosities for the planets in a 3:1 period ratio]{Planetary accretion rates (top row) and masses (bottom row) as a function of time for different $\alpha$-viscosities. Here the planets are fixed in a 3:1 period ratio: the inner planet is shown in red and the outer one in blue. As in \cite{Bergez2020}, the accretion rates slightly increase with increasing viscosity. The oscillations in the accretion rate at low viscosities are due to the presence of vortices. The inner and outer planets display similar accretion rates, leading to similar planet masses.}
        \label{2p:res31_mdot_allalphas}
\end{figure*}

As mentioned previously, the turbulent viscosity is parametrized by the $\alpha$-viscosity parameter. Disk turbulence increases with increasing $\alpha$, leading to faster evolving disks. We show in Fig. \ref{2p:res31_mdot_allalphas} the planetary accretion rates (top row) and the resulting masses (bottom row) for planets in disks with different $\alpha$ parameters. From left to right, $\alpha$ increases from $10^{-4}$ to $10^{-2}$. The behavior of the accretion rates is the same as in \cite{Bergez2020}: reducing the viscosity induces a slightly lower planetary accretion due to a slower flow of gas in the vicinity of the planet.

The difference in accretion rates between the inner (red) and the outer (blue) planet slightly evolves as a function of time and viscosity as shown on the top row of Fig. \ref{2p:res31_mdot_allalphas}. At the beginning, the planets accrete in the first \cite{Machida2010} accretion regime (dominated by Bondi accretion), leading to a larger inner planet. The flip in accretion rate is due to the switch of accretion regime, from a Bondi to a Hill dominated accretion scheme \citep{Machida2010,Bergez2020}. While the inner planet accretes slightly more in the first regime of accretion, the accretion rates become more similar when the planets accrete in the second accretion regime.

This similarity in accretion rates results in planets of similar masses (bottom row of Fig. \ref{2p:res31_mdot_allalphas}). Even if the planets are more massive in the high viscosity case than in low viscosity disks, the differences between the inner and outer planet masses does not seem to be highly influenced by the change of $\alpha$-viscosity. This can be expected as the Machida accretion recipe does not directly depend on this parameter (see Eq. \ref{eq:accroutine}). The influence of the $\alpha$-viscosity is indirect, as it influences the gas flow around the planet, changing the surface density $<\Sigma>_{0.9r_H}$ from which the planets accrete their gas.

At low viscosity, instabilities are triggered \citep{Klahr2003,Fu2014}. The Rossby wave instability \citep[RWI;][]{Lovelace1999,Li2001} produces vortices at the locations of steep pressure gradients, such as produced by the gaps of massive planets. The presence of vortices modifies the flow of gas in the vicinity of the planets, creating the oscillations observed on the top left panel of Fig. \ref{2p:res31_mdot_allalphas}. In the case of two accreting planets, vortices are produced at three different locations: at the outer edge of the outer gap, in between the planets and even at the inner edge of the inner gap. We show in Appendix \ref{appendix_2Dgas_2p} the 2D ($r,\phi$) surface density maps of the disk containing the planets in the 3:1 period ratio at low viscosity ($\alpha=10^{-4}$ and $h=0.05$) at three different times. The strength of these vortices depend on the growing timescale of the planets \citep{Hammer2017,Hammer2021}. However, here, due to the presence of the second planet, the vortex between the planets quickly vanishes (in less than $4\times10^4$ yrs). The strongest vortex (i.e., with the highest over-density) is the one located at the outer edge of the gap of the outer planet and takes about $10^5$ yrs to vanish. 

\subsection{Influence of the aspect ratio}
\label{sec:2p_diffh}

Another way to study the impact of the disk viscosity is to modify the disk's aspect ratio. We show the impact of this parameter in Fig. \ref{2p:res31_mdot_allh}. Here, $\alpha$ is fixed to $10^{-3}$ and $h$ varies from 0.03 (left panel) to 0.07 (right panel). The fiducial value of 0.05 is shown in the middle panel for reference. It should be noted that changing the constant aspect ratio from 0.03 to 0.05 accounts for a reduction of viscosity $\nu$ by a factor of 2.8 while enhancing the aspect ratio from 0.05 to 0.07 increases the disk viscosity by a factor of 2.0.

Even if the change in the disk kinematic viscosity is weaker than when the $\alpha$-viscosity is varied in the previous section (change of a factor of 10), the accretion rates behavior are significantly different depending on the aspect ratio. This arises from the direct dependence of the accretion rate recipe on the aspect ratio, while it is indirectly dependent on $\alpha$. When the gas disk scale height $H$ increases, it requires a larger Hill sphere radius for a planet to switch from the Bondi regulated accretion regime to the Hill regulated one \citep{Machida2010,Bergez2020}. This can be seen on the top row of Fig. \ref{2p:res31_mdot_allh}: the flip in accretion rates occurs at later times for higher aspect ratios. At low aspect ratio (left panel), the planets initial mass is large enough to start accreting immediately in the second regime. In this case, the accretion rates always decrease in time, proportionally to the local surface density.

The timing of the accretion switch between the Bondi and Hill regimes has an important influence on the evolution of the planet masses and their mass ratios. As the inner planet accretes significantly more than the outer one when they are limited to the Bondi accretion regime, a later switch results in more diverse planetary masses. Therefore, the difference between the inner and outer planets increases with an increasing aspect ratio.

\subsection{Impact on the planets' mass ratio}
\label{sec:2p_massratiosdiffnu}

\begin{figure*}[t]
        \centering   
        \includegraphics[width=18cm]{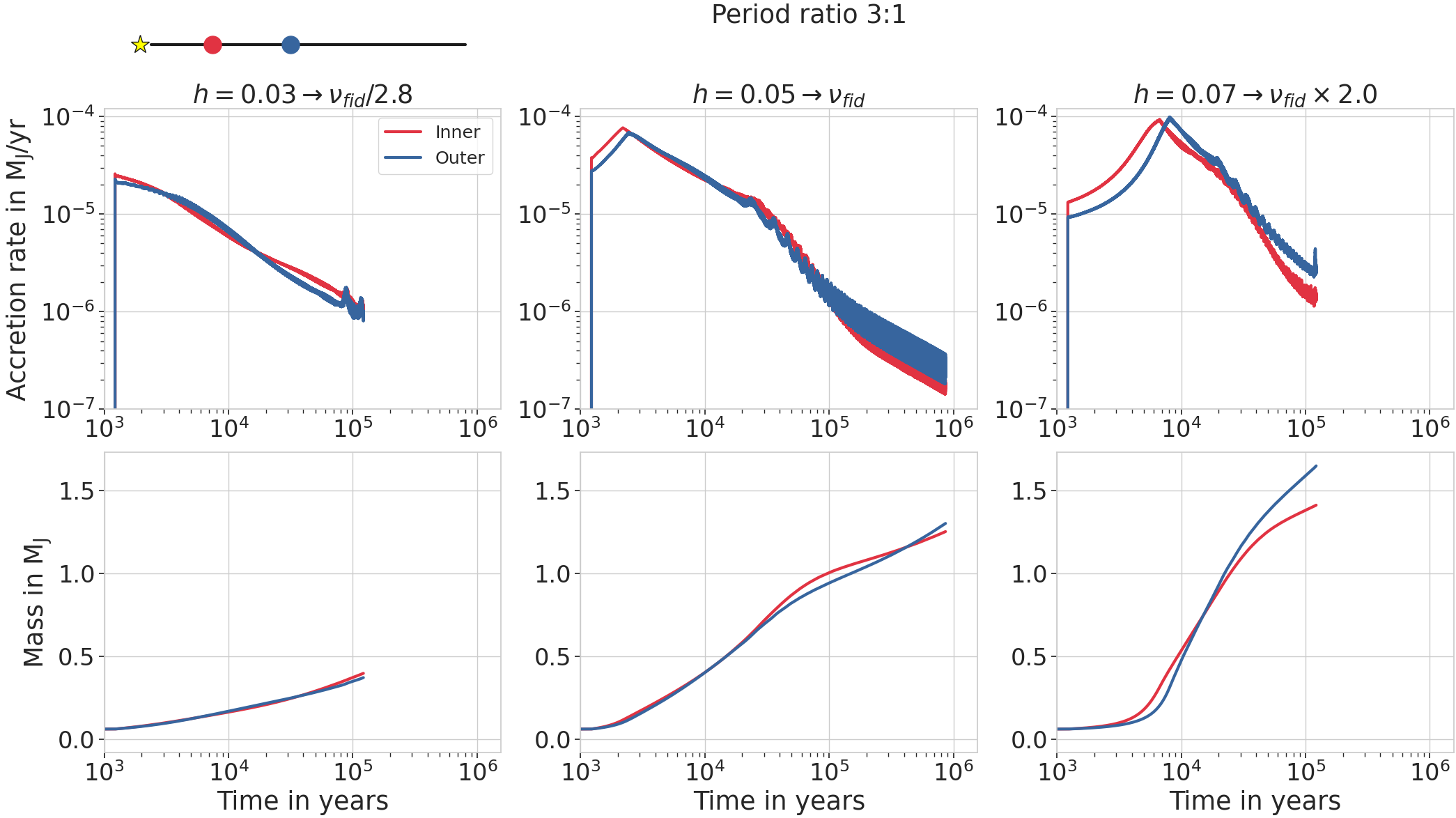}
        \caption[Planetary accretion rates and masses as a function of time for different aspect ratios for the planets in a 3:1 period ratio]{Same as Fig. \ref{2p:res31_mdot_allalphas}, but for different aspect ratios, increasing from left to right, with $\alpha = 10^{-3}$. We note on the top of each row the difference in kinematic viscosity caused by the change in the disk's aspect ratio ($\nu = \alpha h^2 r^2 \Omega_K$). The differences in the planetary accretion rates originate from the structure of the accretion formula itself  (see Eq. \ref{eq:accroutine}). At a low aspect ratio ($h = 0.03$, left panel), the accretion starts immediately in the second Machida regime. On the other hand, for $h = 0.07$, the change of accretion regime occurs at a higher mass and therefore at a later time, leading to a very different planet mass evolution.}
        \label{2p:res31_mdot_allh}
\end{figure*}

\begin{figure*}[t]
        \centering   
        \includegraphics[width=18cm]{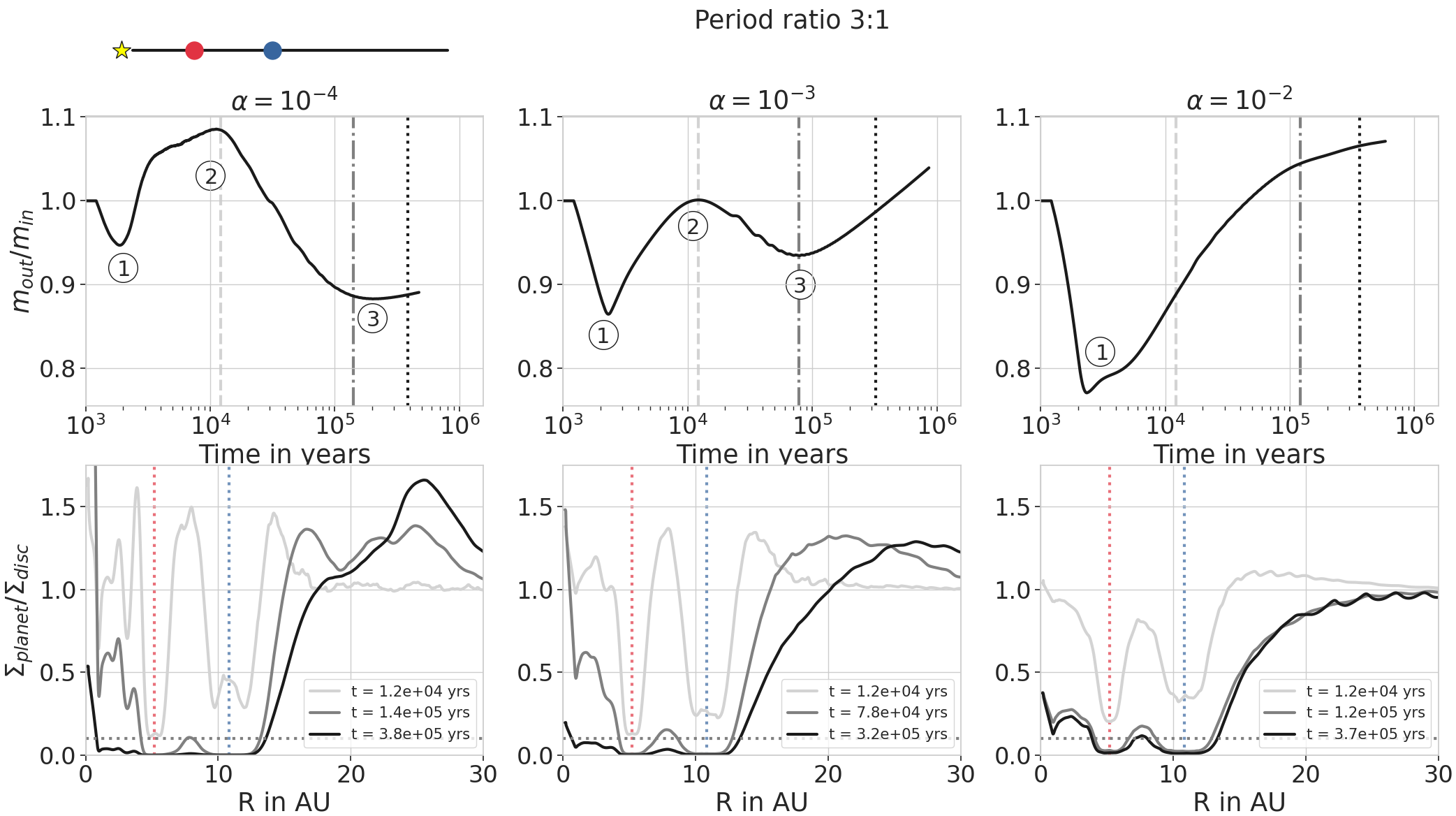}
        \caption[Mass ratio]{Mass ratio (top row) and perturbed surface density profiles (bottom row) as a function of time for different $\alpha$-viscosities. As in Fig. \ref{2p:res31_mdot_allalphas}, the planets are fixed at the 3:1 period ratio positions, represented by the two vertical dotted lines in the bottom row. The horizontal dotted lines in the bottom panels mark the $\Sigma_{planet}/\Sigma_{disk} = 0.1$ threshold, where we consider that a gap is opened \citep{Crida2006}. In the top panels, a decreasing (respectively increasing) mass ratio indicates that the inner (respectively outer) planet is accreting more than the other planet. Different flips are observed in the evolution of the mass ratios, marked in circled numbers. The color of the surface density profiles shown in the bottom row corresponds to the color of the different vertical lines in the top row and represents different times. The last snapshot of the perturbed surface density is taken $2.4\times10^5$ yrs (20000 orbits of the inner planet) after the last mass ratio flip, shown by the vertical dotted black line. The first flip \circled{1} can be explained by the accretion formula itself (see Eq. \ref{eq:accroutine}), while the other two flips correspond to the evolution of the surface density.}
        \label{2p:res31_mratio_allalphas}
\end{figure*}

\begin{figure*}[t]
        \centering   
        \includegraphics[width=18cm]{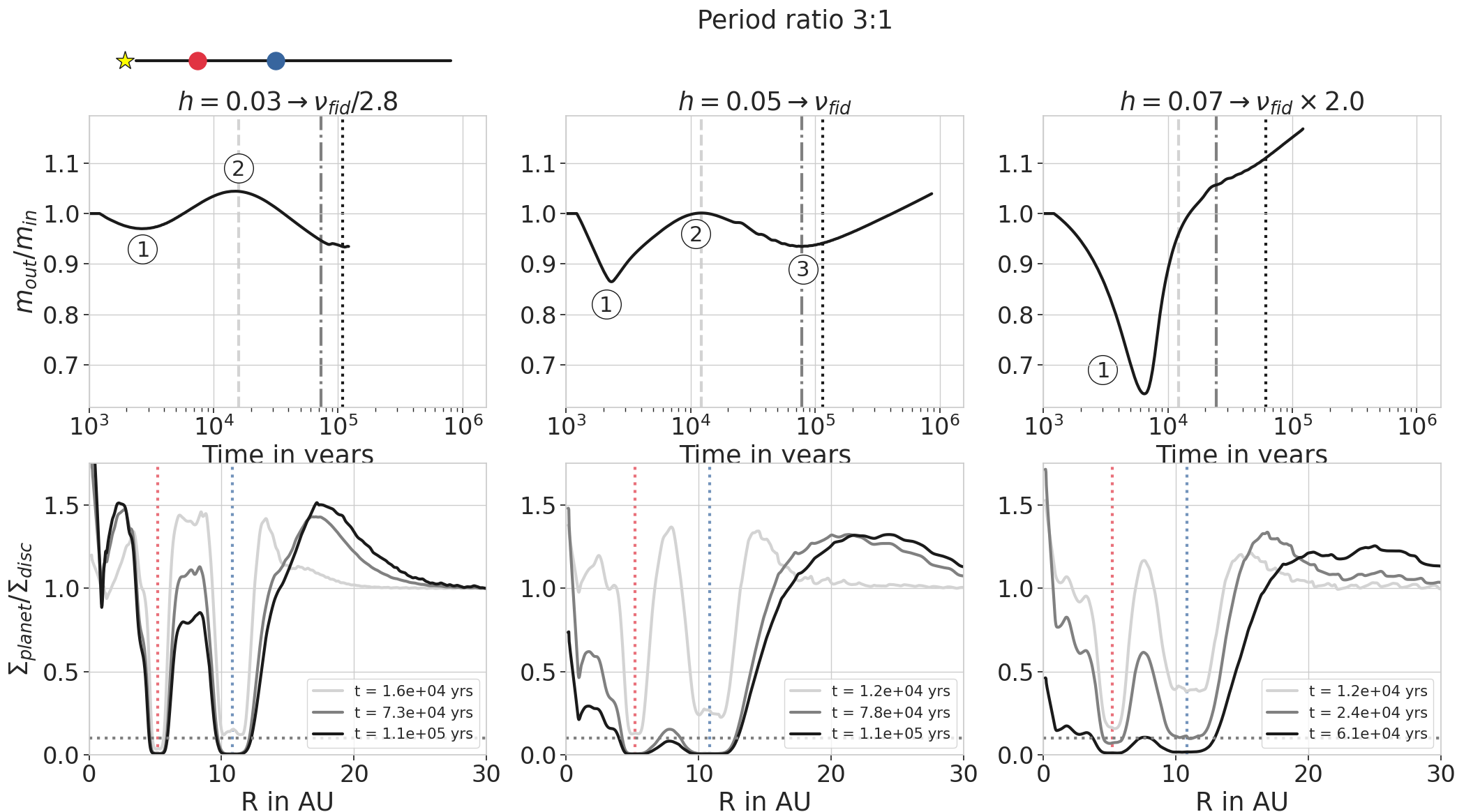}
        \caption[Same as Fig. \ref{2p:res31_mratio_allalphas} but for different aspect ratios]{Same as Fig. \ref{2p:res31_mratio_allalphas} but for different aspect ratios, with $\alpha = 10^{-3}$. As in Fig. \ref{2p:res31_mdot_allh}, we note at the top of each row the difference in kinematic viscosity caused by the change in the disk's aspect ratio. Here the mass ratio evolutions differ in amplitude, but they present the same behavior as in Fig. \ref{2p:res31_mratio_allalphas}. The difference in amplitude originates from the direct dependence of the accretion routine on the aspect ratio, and not on the $\alpha$-viscosity (see Eq. \ref{eq:accroutine}).}
        \label{2p:res31_mratio_allh}
\end{figure*}

In order to better understand the differences in planetary mass, we analyze in Fig. \ref{2p:res31_mratio_allalphas} and Fig. \ref{2p:res31_mratio_allh} the ratios of the planetary masses. We focus first in Fig. \ref{2p:res31_mratio_allalphas} on the impact of the $\alpha$-parameter on these ratios. The top row represents the mass ratio as a function of time. We arbitrarily decided to show the ratio of the outer planet mass divided by the inner planet mass: this means that when the ratio is decreasing, the inner planet accretes more than the outer one and vice versa. Depending on the viscosity, we see that the ratio shows different flips in time. A first flip \circled{1} occurs at each viscosity around $2.5\times 10^{3}$ yrs. This flip originates from the accretion rate switch discussed in the previous paragraphs: the planets become massive enough to change their accretion regime, resulting in a higher accretion rate for the outer planet (increasing mass ratio). 

A second flip \circled{2} is observed around $10^{4}$ yrs, at all viscosity except for $\alpha = 10^{-2}$ (right panel, which is discussed below). This flip is linked to the formation of deep planetary gaps. In the bottom panels of Fig. \ref{2p:res31_mratio_allalphas}, we show the perturbed surface densities at the different times marked by the vertical lines in the top panel. The perturbed surface density is defined as the surface density of the disk in presence of the planets normalized to the surface density of a disk without planets at the same time ($\Sigma_{perturbed} (t) = \Sigma_{planet} (t)/\Sigma_{disk} (t)$). These perturbed surface density profiles are used to determine when a gap is opened. We use the definition suggested by \cite{Crida2006}: a gap is considered opened when the gas surface density is depleted by $90\%$ compared to a disk without planets. This threshold is represented by the horizontal gray dotted line at $\Sigma_{planet}/\Sigma_{disk} = 0.1$. The second flip \circled{2} in the mass ratio occurs when the inner planet reaches this threshold: when the inner planet opens a deep gap, it starts to accrete more than the outer planet.

This link between the gap opening of the inner planet and the decrease of the mass ratio can be explained by the impact of gas accretion on gap opening described in \cite{Bergez2020}. At low viscosity, gas accretion does not help carve a gap because the disk reaction time is long. The gas is therefore dominantly pushed away from the planets orbit by gap opening, enhancing the surface density in between the planets as well as in the inner and outer regions of the disk. This results in a perturbed surface density larger than one (light gray line in the bottom panels of Fig. \ref{2p:res31_mratio_allalphas}). When the gap is opened, the planetary gas accretion becomes the dominant process influencing the gas distribution. As the inner planet opens its gap first, it starts by depleting the material present in between the planets and in the inner region of the disk, leading to an inner planet with a higher accretion rate than the outer one. Then, the outer planet also opens a deep gap, helping the inner planet deplete the material located in between them. At this stage, the amount of gas present around the inner planet (at $r < r_{p,out}$) is dictated by three different processes. Gas is removed from this region of the disk by (i) the accretion onto the planets and (ii) the accretion onto the star, and is replenished by (iii) the viscous diffusion of the gas from the outer part of the disk through both planet gaps. At low viscosity, only a small amount of gas manages to diffuse through the gaps of both planets. This results in a depletion in gas of the inner disk, resulting in the starvation of the inner planet.

The third flip \circled{3} in the mass ratio, shown in the top panels of Fig. \ref{2p:res31_mratio_allalphas}, corresponds to the moment when the outer planet starts accreting more than the starved inner planet, because it is supplied in gas by the outer region of the disk. The depletion of the inner disk can be seen in the corresponding perturbed surface density profiles of the bottom row of Fig. \ref{2p:res31_mratio_allalphas}: the gray line shows a depleted region in between the planets and in the inner disk. We additionally show the surface density profiles at 20 000 inner planetary orbits ($\simeq 2.4 \times 10^5$ yrs) after the last mass ratio flip \circled{3}: after this time, the inner disk is almost completely emptied in gas, with a perturbed surface density of less than 0.1 within 12 AU. It is clear that after this time, the outer planet will accrete more than the inner one, until it becomes the most massive planet of the system. The only differences between the $\alpha = 10^{-4}$ (left panel) and $\alpha = 10^{-3}$ (middle panel) cases are the delay in time of the flips due to different viscous timescales and the presence of vortices at $\alpha = 10^{-4}$, influencing the accretion rates of both planets, as mentioned earlier.

The behavior of the mass ratio in a high viscosity disk (right panel) is slightly different than at lower viscosities. As shown in \cite{Bergez2020}, at high viscosity, gas accretion helps gap formation. Therefore, the material located in between the planets and the inner disk is immediately depleted by gas accretion and viscous stellar accretion. As these two regions are not enhanced in gas by gap opening, the inner planet accretion rate slowly reduces as the inner disk is immediately depleted in gas, meaning that no additional mass ratio flip is observed except for the very first one originating from the accretion recipe. At this viscosity, the gas manages to diffuse efficiently through the gaps of both planets, avoiding a complete depletion of the inner region: this can be seen by comparing the surface density profiles at two different times toward the end of the simulation (marked by the gray and black lines on the right panels). Indeed, the difference in the profiles after 2 000 orbits of the inner planet is small, meaning that the disk gas flow is high enough to prevent the total depletion of the inner region, unlike at lower viscosities (middle and left panels). However, in this configuration, even if the material around the inner planet is replenished by viscous evolution, the amount of gas diffused through both gaps is not high enough to allow the inner planet to accrete more than the outer planet.

These behaviors at high and low viscosity are also observed when we vary the disk aspect ratio. In Fig. \ref{2p:res31_mratio_allh}, we show the evolution of the mass ratios as a function of time (top row) as well as the perturbed surface density profiles at given times (bottom row) for the different aspect ratios presented in Fig. \ref{2p:res31_mdot_allh}. As the behaviors described in the previous paragraphs are only dependent on the gas kinematic viscosity $\nu$, we recover the same behaviors when we change the aspect ratio: at low viscosity, the mass ratios are highly influenced by gap formation whereas at high viscosity, the mass ratio does not present more than one flip. However, due to the dependence of the accretion recipe on the aspect ratio, the amplitudes of the mass ratios are highly dependent on $h$. While the mass ratios reached values between 0.8 and 1.1 for the different $\alpha$ parameters (Fig. \ref{2p:res31_mratio_allalphas}), here the planets show a larger spread in mass ratio at the beginning of the simulations in disks with larger aspect ratios. However, once the gaps are formed in all cases and the inner disk is depleted, all the mass ratios stay quite close to 1, with an outer planet less than 1.2 times more massive than the inner planet.

As a conclusion, the mass ratio between the planets is highly and mainly dependent on the disk kinematic viscosity $\nu$. The resulting mass ratios are close to 1 even after $0.5$ Myrs of evolution, leading to planets with rather similar masses. In all the explored cases, we expect the outer planet to become the most massive planet of the system. 

\section{Single accreting planet compared to two accreting planets}
\label{sec:2p_2p1pcomp}

The presence of the second planet highly influences the growth of the inner planet. In order to quantify the effect of the presence of the second planet, we compare the growth of the two planets to two separate simulations where the planets are alone in their disk.

\subsection{Accretion rate and mass comparisons}
\label{sec:2p_2p1p_mdotmass}

\begin{figure*}[t]
        \centering   
        \includegraphics[width=18cm]{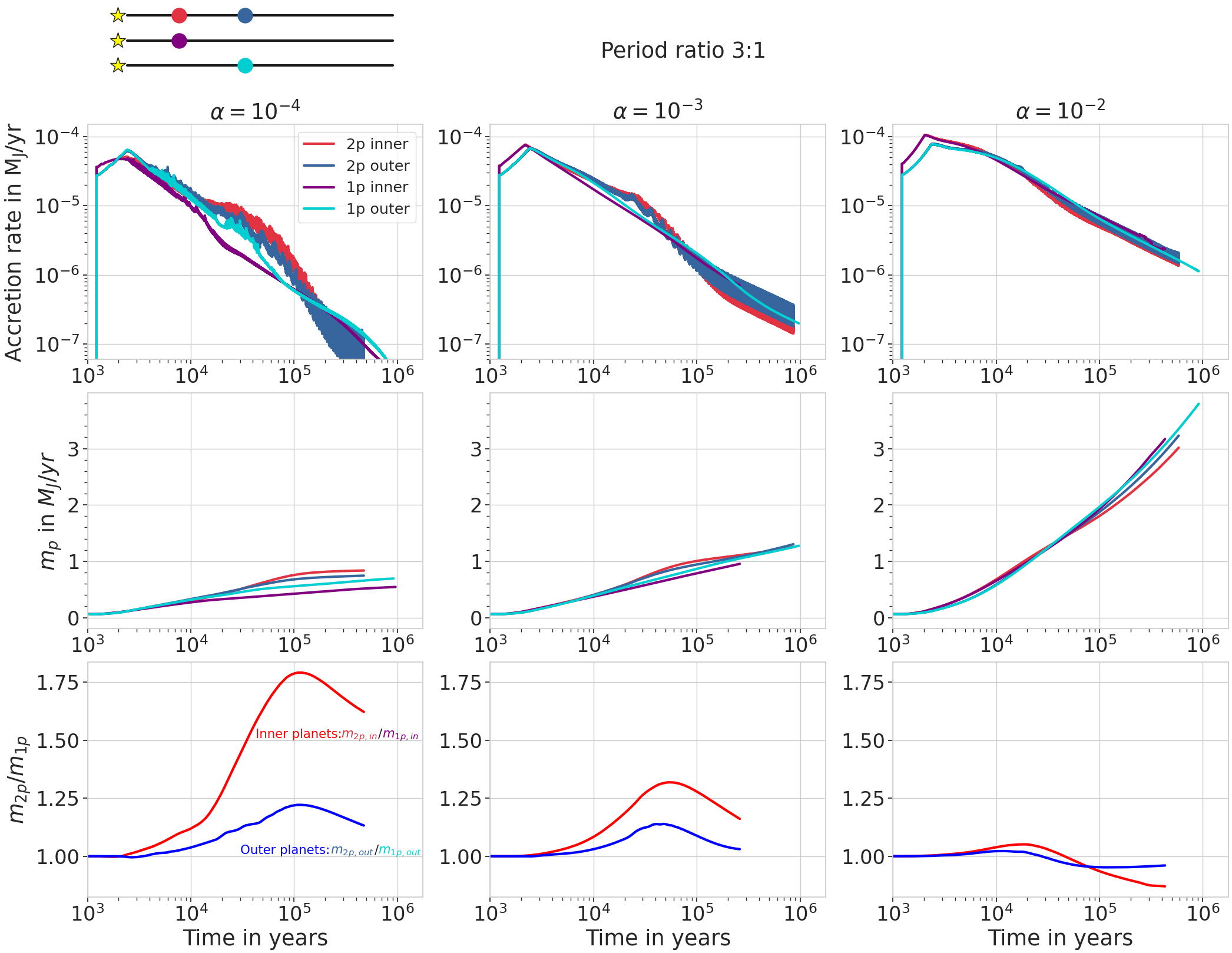}
        \caption[Comparison between single accreting planets and two simultaneously accreting planets.]{Comparison between single accreting planets and two simultaneously accreting planets. The planets are fixed at positions corresponding to the 3:1 period ratio. As in Figs. \ref{2p:res31_mdot_allalphas} and \ref{2p:res31_mdot_allh}, the top row presents the accretion rates as a function of time and the middle row the planet masses. In the bottom row, we show the ratio of the masses in the single- and two-planet case: the red line represents the ratio of the inner planets ($m_{2p,in}/m_{1p,in}$), and the blue line represents the outer planets' ratio ($m_{2p,out}/m_{1p,out}$). At low viscosity (left panels), the differences with the single-planet case originate from the additional formation of vortices in between the planets, which enhances the accretion rates during the vortex lifetimes. At high viscosity (right panel), while the outer planet is slightly impacted by the presence of the inner planet, we observe the starvation of the inner planet. We expect to see similar behavior at lower viscosities but delayed in time, due to longer viscosity timescales.}
        \label{2p:2p1p_mdot_mass}
\end{figure*}

It has been shown in Sect. \ref{sec:2p_massratiosdiffnu} that the outer planet has the capacity to starve the inner planet once the gaps are formed and the inner disk is depleted. Consequently, the differences between two accreting planets and single accreting planets should increase with time. We show in Fig. \ref{2p:2p1p_mdot_mass} the comparison between three different simulations: the first one, represented by the blue and red colors, considers the simultaneous accretion of two planets in the 3:1 period ratio as in the previous section; in the second simulation, the disk hosts only a single planet located at the position of the inner planet (purple line); in the third simulation, the disk hosts also a single planet located this time at the position of the outer planet (cyan line). Each planet configuration is represented in the top left corner of the figure. As in Sect. \ref{sec:2p_diffalphas}, different $\alpha$ are shown in the different columns, increasing from left to right.

The different planetary gas accretion rates are shown in the top row of Fig. \ref{2p:2p1p_mdot_mass}. At high viscosity, in the right panel, the accretion rate between the single planets and the two planets are very similar. As expected, only the accretion rate of the inner planet is significantly impacted by the presence of the second planet: after $2\times10^4$ yrs, the accretion rate of the inner planet in the two-planet case (red line) starts to be reduced compared to the single-planet case (purple line). This is particularly visible in the bottom right panel of Fig. \ref{2p:2p1p_mdot_mass}, where we compare the masses in the two-planet case to the single-planet case: the red line represents the mass ratio of the inner planets ($m_{2p,in}/m_{1p,in}$) while the blue line represents the mass ratio of the outer planets ($m_{2p,out}/m_{1p,out}$). After $2\times10^4$ yrs, the red line continuously decreases, meaning that the planet at the inner position has a reduced accretion rate in the two-planet case compared to the single planet. This is due to its starvation, as discussed in the previous section. 

At high viscosity, the outer planet is only slightly impacted by the presence of the inner planet. While their accretion rate seems very similar (blue and cyan line in the top right panel), the mass ratio shows a slight reduction of the mass in the two-planet case (blue line slightly below one in the bottom right panel). This originates also from the depletion of the inner disk: the gas accretion rate of the outer planet only relies on the material located outside the planet's orbit in the two-planet case, while the single planet accretes material from both the outer and inner disk. 

At lower viscosities, the impact of the presence of the second planet occurs earlier and is more significant. Focusing on the intermediate viscosity ($\alpha=10^{-3}$, in the middle column), we see that the accretion rates (top panel) in the two-planet case are enhanced after $\sim 10^4$ yrs for both the inner and the outer planet compared to the single planets from this point in time. This enhancement, absent at high viscosity, originates from the gap opening process: as mentioned in Sect. \ref{sec:2p_massratiosdiffnu}, at this viscosity, gap opening pushes material away from the vicinity of each planet, enhancing the surface density around them. In this case, the planets are close enough to each other to push material in the feeding zone of the neighboring planet. Therefore, the inner planet pushes material toward the outer planet and vice versa. Each accretion rate is enhanced until the depletion of the inner disk ($r<r_{p,out}$), resulting in lower accretion rates than in the single-planet cases. This enhancement can also be seen in the evolution of the planet masses (bottom and middle panels of Fig. \ref{2p:2p1p_mdot_mass} for $\alpha=10^{-3}$). In the long term, the evolution of the mass ratios between the two-planet and single-planet cases are expected to behave like at high viscosity: the inner planet will be starved in gas, leading to a smaller inner planet in the two-planet case compared to the single-planet case. As for the outer planet, once the inner disk is depleted, it is fed only by the outer disk in the two-planet case whereas the single outer planet accretes also from the inner disk. This leads to a decreasing outer planet mass ratio $m_{2p,out}/m_{1p,out}$ until reaching an almost constant ratio, similarly to the high viscosity case.

Regarding the lowest viscosity case ($\alpha = 10^{-4}$, in the left column of Fig. \ref{2p:2p1p_mdot_mass}), the presence of additional vortices located in between the planets significantly alters the accretion rates of the two accreting planets compared to the single planets. Even if the single planets also produce vortices, leading to the oscillations observed in their accretion rate too, the additional vortex present in between the planets quickly enhances the accretion of both planets. Except from this non-negligible influence, the overall evolution of the planet masses follows the trend observed at $\alpha = 10^{-3}$, where we expect the different flips to occur at later times due to the larger viscous timescale.

\begin{figure*}[t]
        \centering   
        \includegraphics[width=18cm]{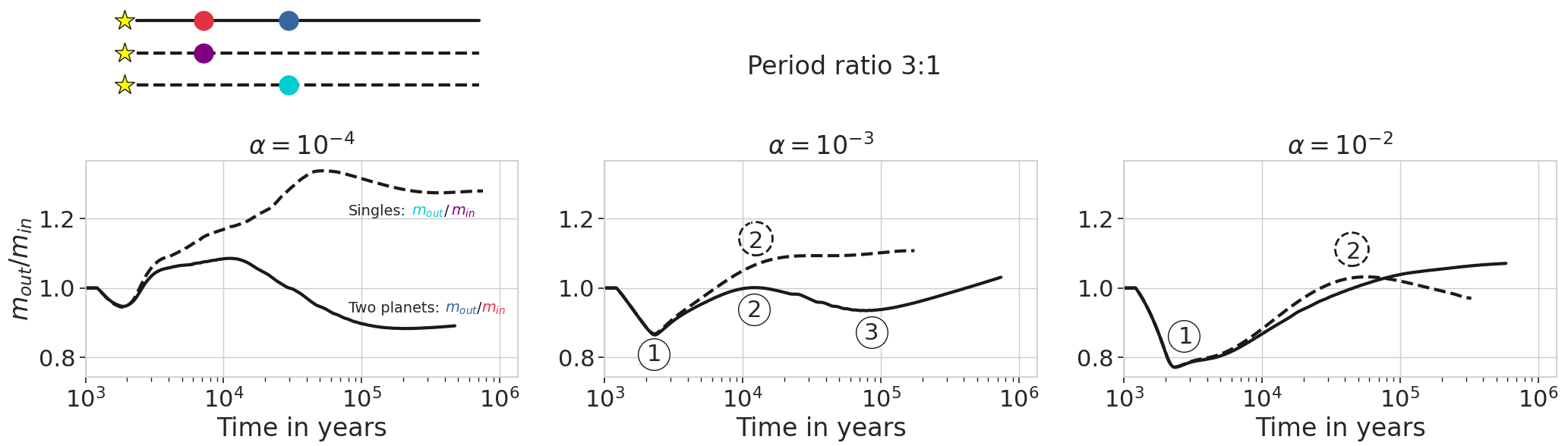}
        \caption[Comparison of the mass ratios in the single planet and two accreting planets cases as a function of time and for different viscosities.]{Comparison of the mass ratios in the single-planet and two-planet cases as a function of time and for different viscosities. Again, the planets are fixed at the 3:1 period ratio positions, as shown in the upper-left corner. In solid lines we show the mass ratios of the outer over the inner planet in the two-planet case, whereas the dashed lines represent the mass ratio of the single planets. Different flips are marked by circled numbers and can be explained by the impact that gap opening and depletion of the inner disk have on the accretion behavior of the planets. }
        \label{2p:2p1p_mratio}
\end{figure*}

%We observe different behaviors when a second planet is added: at high viscosity, the inner planet in the disk with two planets is quickly starved by the outer one, resulting in a single flip \circled{1}. The mass ratio in the single planets case present a second flip \circled{2} because the flow of gas through a single planetary gap is high enough to maintain a higher accretion of the inner single planet compared to the single outer planet. In the two-planet case, the gas flow to the inner disk is too reduced by the presence of the two planets, resulting in a larger outer planet (increasing mass ratio). At lower viscosities, the single planet case does not present the flips \circled{2} and \circled{3} originating from the material pushed by gap opening of the neighbouring planet and depletion of inner disk ($r<r_{p,out}$) respectively.

Finally, we also compare the masses of the outer planets with the inner planets, as in Sect. \ref{sec:2p_massratiosdiffnu}. At intermediate viscosity ($\alpha=10^{-3}$, middle panel of Fig. \ref{2p:2p1p_mratio}), the mass ratio in the single-planet cases (dashed line) features only one flip \circled{1} before reaching a plateau after $\sim2\times10^4$ yrs. As before, the first flip \circled{1} is due to the accretion recipe, and it occurs at the same time in the two-planet case as in the single-planet cases. The second flip \circled{2} in the case of the two planets in the same disk, due to gap opening, is absent in the single-planet cases. This is expected because it is the presence of the second planet creating a gap that enhances the accretion of the inner planet, which is not the case in the single-planet simulations. With our disk profile (a surface density power law of -1 and constant aspect ratio), the single planet located at the outer position accretes naturally more than the inner planet, leading to an increasing mass ratio. The plateau observed after $\sim2\times10^4$ yrs corresponds to the moment when the single planets open a deep gap: then, the accretion of each planet is mainly governed by the flow of gas originating from the outer region of the disk. With our disk profile and due to the close proximity of the planets, the gas flow from outside of the inner planet position is similar to the flow from outside of the outer planet position, leading to similar accretion rates. The same behavior is observed at low viscosity ($\alpha = 10^{-4}$, in the left panel of Fig. \ref{2p:2p1p_mratio}), considering the perturbations and enhancement produced by the vortices.

At high viscosity ($\alpha=10^{-2}$, right panel of Fig. \ref{2p:2p1p_mratio}), the behavior of the single-planet simulations is again different compared to the simulations with two planets. The absence of additional flips after $2\times10^3$ yrs in the two-planet case is explained by the relative absence of impact of gap opening at this high viscosity (see Sect. \ref{sec:2p_massratiosdiffnu}): the viscosity is high enough to cause the immediate depletion of the inner disk via both viscous accretion toward the star and the accretion of the inner planet, immediately leading to its starvation. However, in the mass ratio of the single planets, we observe another flip \circled{2} around $4\times10^4$ yrs. After this time, the inner planet accretes more than the outer one. This originates from the flow of gas reaching the inner disk once each gap is opened: at this viscosity, the gas can significantly flow through the gaps of the planets. However, it is easier to flow through the gap of the single planet located in the inner region as it is less wide than the  gap of the single planet located in the outer region. This results in a inner disk that is more depleted in the case of the outer single planet than in the case of the inner single planet, leading to the reduction of the accretion rate onto the outer single planet.

\subsection{Impact on the accretion onto the star}
\label{sec:2p_stellaracc}

\begin{figure*}[t]
        \centering   
        \includegraphics[width=18cm]{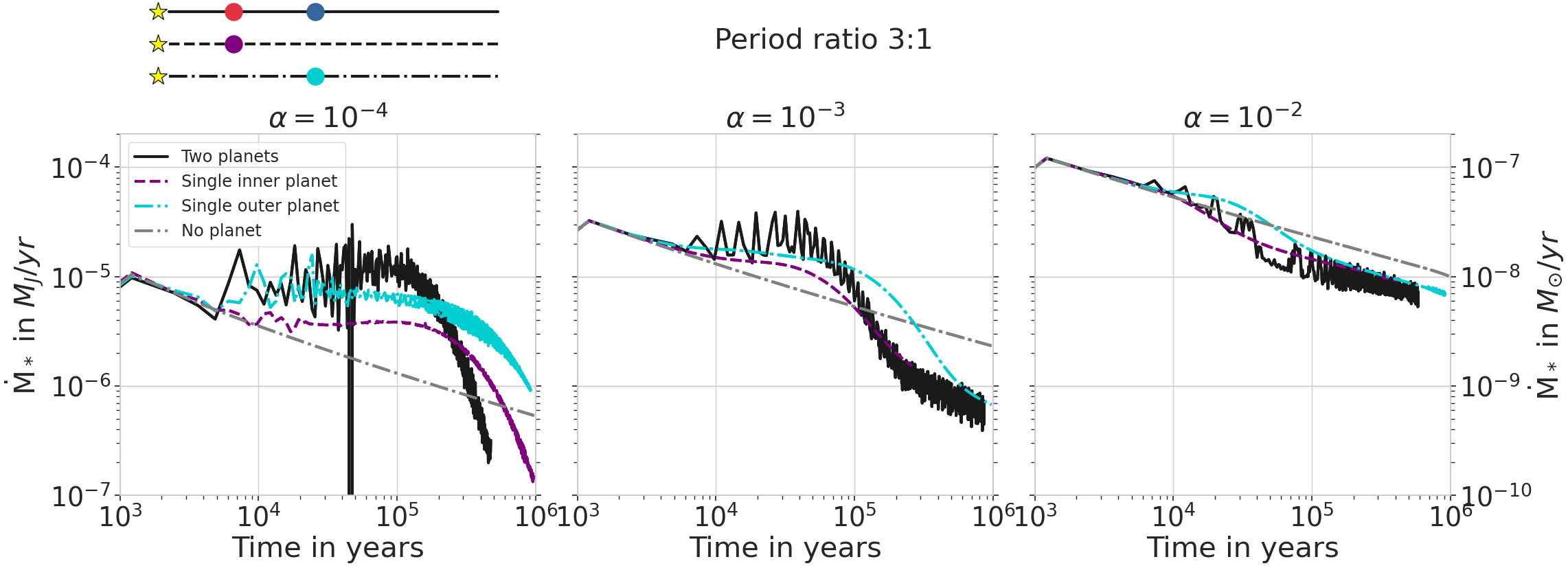}
        \caption[Influence of a second accreting planet on the stellar accretion rate at the inner edge of the disk (0.2 AU).]{Influence of a second accreting planet on the stellar accretion rate at the inner edge of the disk (0.2 AU). As in Figs. \ref{2p:2p1p_mdot_mass} and \ref{2p:2p1p_mratio}, the planets are located in the 3:1 period ratio positions, with an increasing $\alpha$-viscosity from left to right. The comparison is made between the two-planet case (solid black line), the single-planet cases (dashed purple line for the single inner planet and dotted cyan dashed line for the single outer planet) and a disk without planets (dotted dashed gray line). The oscillations present in the case of the two accreting planets are due to the overlap of their spiral arms, coupled with the presence of vortices at low viscosity. This sometimes results in a negative radial velocity, explaining the dip seen at $\alpha = 10^{-4}$ due to the logarithmic scale. Compared to a disk with no planet, the accretion onto the star is only reduced by up to a factor of 3 in the presence of multiple giant planets, similar to the scenario with a single planet \citep{Bergez2020}.}
        \label{2p:2p1p_starmdot}
\end{figure*}

The presence of a single gap opening planet can alter the gas accretion onto the central star \cite[e.g.,][]{Manara2019,Bergez2020}. Here, we investigate the impact of the presence of a second planet on the evolution of the stellar gas accretion. The stellar gas accretion rate is defined by the flow of mass through the inner edge of the disk: $  \dot{M}_* = -2\pi \; r_{in} v_{r,in} \Sigma_{in},$ where $  v_{r,in}$ and $  \Sigma_{in}$ are the radial velocity and surface density at the inner boundary located at $r_{in} = 0.2$ AU. In Fig. \ref{2p:2p1p_starmdot}, we compare the stellar accretion rates of disks that contain two accreting planets in the 3:1 period ratio (solid black line), disks that contain single planets located at the inner (purple dashed line) and outer (cyan dotted dashed line) positions of the two-planet simulation, or disks that host no planets at all (gray dotted dashed line). Again, the $\alpha$-viscosity increases from left to right. Independent of the viscosity, the stellar accretion rate in the presence of two planets features oscillations: they originate from the periodic overlap of the planets spiral density arms, locally enhancing the surface density of the gas.

At intermediate viscosity ($\alpha = 10^{-3}$, middle panel of Fig. \ref{2p:2p1p_starmdot}), the presence of the planets has two distinct impacts. Before $10^4$ yrs, the planets are not large enough to influence the accretion onto the star. Between $10^4$ yrs and $10^5$ yrs, the disks hosting planets harbor an enhanced stellar accretion rate compared to the disk without planets. This originates from gap opening: material is pushed to the inner region, feeding the star. After $10^5$ yrs, planetary gas accretion and gaps prevent part of the gas from reaching the inner region, leading to the decrease in the stellar accretion rate. It takes more time for the disk hosting the single planet located at the outer position to reduce the stellar accretion rate due to the larger inner disk present in this configuration. After a given time (here after $2\times10^5$ yrs), the flow of gas through the gaps reach a quasi equilibrium state, leading to a linear evolution of the stellar accretion rate, with a similar slope compared to the slope of the stellar accretion rate of the disk without planets. At this stage, the stellar accretion rate is reduced by a factor of between 4 and 5 when the disk hosts two accreting planets compared to the disk without planets.

The enhancement produced by the two accreting planets is slightly more pronounced than in the single-planet case as it includes the material pushed by both planets. As one could expect, the decrease of the stellar accretion rate in the two-planet case occurs at the same time as in the disk hosting the single planet located at the inner position. However, at this viscosity, the reduction of the stellar accretion rate is only barely influenced by the presence of the second planet, resulting in a decrease of only $\sim30\%$ compared to the single inner planet case. This means that the flow of gas reaching the inner disk is mostly governed by the influence of the inner planet.

At high viscosity, the viscous spreading of the gas prevents the enhancement of the stellar accretion rate: the material pushed to the inner regions by the gap opening planets is both less important than at lower viscosity and quickly diffused toward the star by viscous spreading. This results in a reduction of the stellar accretion rate after $10^4$ yrs only. The gas is efficiently diffused through the planet gaps, maintaining a high stellar accretion rate, even in presence of multiple accreting planets. Here, the stellar accretion rate is only reduced by up to a factor of 2.5 compared to the accretion in a disk without planets. Moreover, the presence of the second planet only influences the stellar accretion rate by less than $20\%$ compared to the cases with single planets. Focusing on the single-planet cases, we see that the stellar accretion reaches a similar equilibrium in both cases, independent of the planet location. As expected, with this high viscosity, the presence of accreting planets barely influences the gas disk evolution.

The opposite behavior is observed at low viscosity ($\alpha = 10^{-4}$, left panel of Fig. \ref{2p:2p1p_starmdot}). Here, the stellar accretion rate is highly influenced by the vortices and by the gaps formed by the planets. Material is efficiently pushed in the inner regions, enhancing the stellar accretion rate by up to a factor of 10 in the disk hosting two planets compared to the accretion of a disk without planets. The reduction of the accretion occurs at later time than at higher viscosity, which is expected because the planets grow slower and gap opening takes more time \cite{Bergez2020}. However, the presence of the second planet in the disk highly influences the stellar accretion rate, because the viscosity is not high enough to diffuse gas through both gaps. Due to the long viscous timescale, we are not able to determine the final reduction factor compared to the higher viscosity cases. At the end of the simulation containing the two planets, the stellar accretion rate is reduced by a factor of 3 compared to the disk without planets, giving a lower estimation of the reduction factor. In Sect. \ref{sec:2p_discuss_ontostar}, we discuss the impact that this reduction factor can have on observations and compare it to other studies.

\section{Influence of the planet separation}
\label{sec:2p_diffseparation}

Our study has focused so far on planets placed at positions corresponding to the 3:1 period ratio. However, the separation between forming planets is not fixed in time as they can dynamically interact with the disk and with each other \citep[e.g.,][]{Baruteau2014,Crida2017}. While we will investigate the impact of the radial evolution of the planets on their growth in a future study, we study in the following section the impact of the planet separation in their growth by placing the planets at different period ratios. 

\begin{figure*}[t]
        \centering   
        \includegraphics[width=18cm]{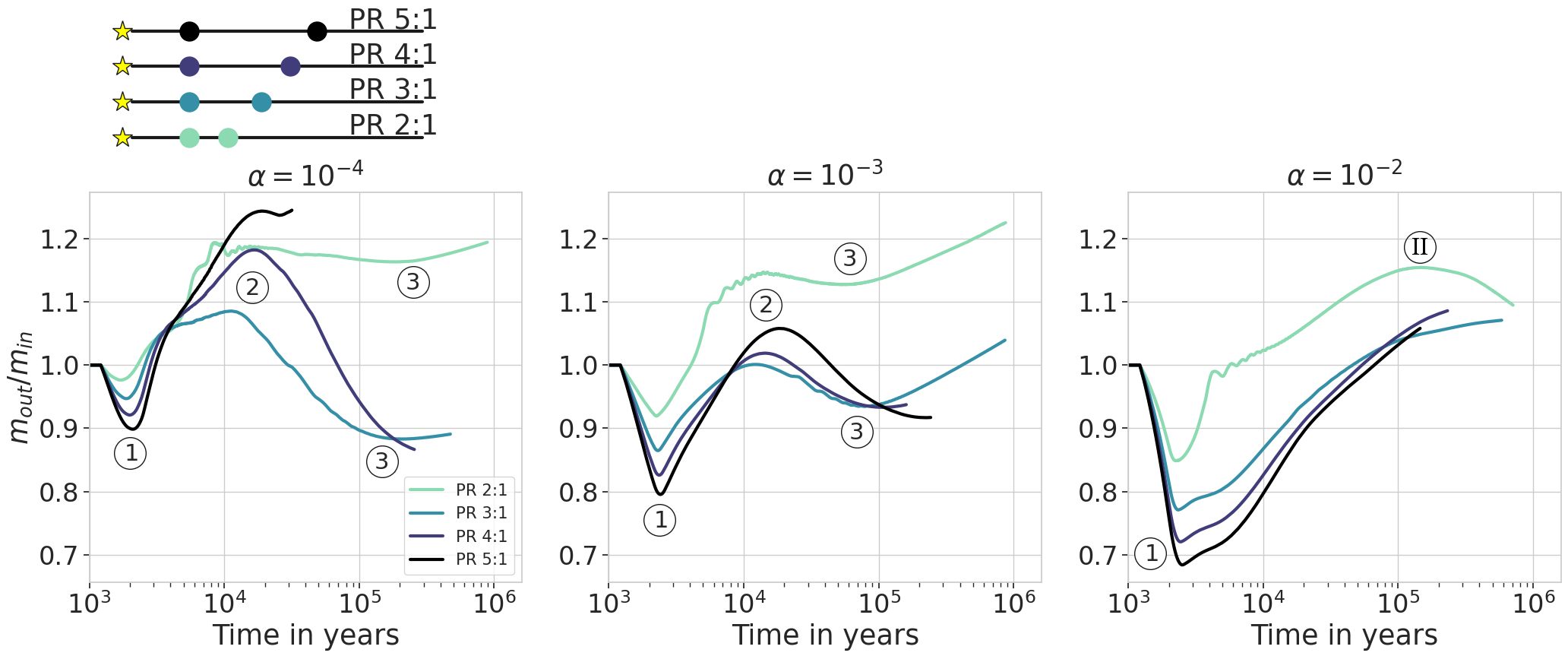}
        \caption[Mass ratio as a function of time for different $\alpha$-viscosities and different period ratios.]{Mass ratio as a function of time for different $\alpha$-viscosities and different period ratios. The darker the line, the farther away the planets are located from each other. The separation between the planets has a small impact on the range of their mass ratio, meaning that planets that start accreting at the same time will have similar masses. The behavior of the mass ratio evolution is not impacted by the planet separation: at all viscosities the mass ratio flips are due to the same reasons described in Fig. \ref{2p:res31_mratio_allalphas}. The exception is for the 2:1 period ratio at high viscosity (right panel): here the planets are close enough and the viscosity is sufficiently high to maintain a significant flow toward the inner disk through the gaps, preventing the inner planet from being starved. This results in an additional mass ratio flip, \circled{II}, similar to the high viscosity single-planet case of Fig. \ref{2p:2p1p_mratio}.}
        \label{2p:allconf_mratio_allalphas}
\end{figure*}

\subsection{Impact on the mass ratio}
\label{sec:2p_mratio_allconf}

As shown in Sect. \ref{sec:2p_massratiosdiffnu}, the planet mass ratio evolution reflects the accretion history of the planets. We show in Fig. \ref{2p:allconf_mratio_allalphas} the mass ratio evolutions of two simultaneously accreting planets located at different period ratios, ranging from 2:1 up to 5:1. These period ratios, represented to scale in the top left corner of the figure, were chosen such that the planets can be considered dynamically stable during their growth, as we neglect their dynamical interactions. Their stability is monitored thanks to their mutual Hill radii \citep{Chambers1996}. As in Fig. \ref{2p:res31_mratio_allalphas}, the mass ratio presented is the ratio of the outer planet mass divided by the inner planet mass and each panel represents an $\alpha$-viscosity, increasing from left to right. A darker line depicts a larger planet separation. 

Independent of the viscosity, we first observe that the first mass ratio flip \circled{1}, originating from the switch of accretion regime in our accretion recipe, is delayed in time when the planets are farther away from each other. Indeed, with our disk profile, a planet located farther away in the disk has a slightly lower gas accretion rate, meaning that it will need more time to reach the mass needed to switch from the Bondi to the Hill accretion regime. This delay affects the similarities between the planets: the farther the planets are from each other, the more extreme are their mass ratios at the beginning of the simulations (i.e., the deeper the first mass ratio flip \circled{1} is). 

Focusing on the high viscosity behavior ($\alpha=10^{-2}$, right panel of Fig. \ref{2p:allconf_mratio_allalphas}), we observe the same behavior as described in Sect. \ref{sec:2p_massratiosdiffnu} except for the 2:1 period ratio. While the other period ratios feature only the first flip \circled{1} due to the rapid depletion of the inner disk via viscous stellar accretion, a second flip \circled{II} is observed at around $10^5$ yrs in the mass ratio of the 2:1 period ratio case. In this configuration, the planets are close enough from each other to facilitate the diffusion of gas through both gaps compared to the other period ratios, as they quickly form a common gap. As the flow of gas to the inner disk is higher, more material is present around the inner planet. In this case, the amount of gas diffusing through both gaps, is high enough to prevent the starvation of the inner planet, leading to a decreasing mass ratio. 

The behavior of the mass ratio at $\alpha=10^{-3}$ (middle panel of Fig. \ref{2p:allconf_mratio_allalphas}) is also the same for each period ratio. The second mass ratio flip \circled{2}, occurring at around $10^4$ yrs, is always due to the formation of the planetary gaps. The delay in the flip originates from the time needed for the outer planet to also significantly enhance the surface density in between the planets via gap opening. The amplitude of the mass ratio between the second \circled{2} and third flip \circled{3} (i.e., the difference between the local maximum \circled{2} and the local minimum \circled{3}) is decreasing with decreasing planet separation. As this moment corresponds to the moment when the inner planet is accreting more thanks to the enhancement of material in between the planets and in the inner disk after gaps are formed, closer planets have less material in between them by construction, leading to a smaller mass ratio amplitude. We note that it also takes more time for planets located farther away from each other to empty the material located between the planets, meaning that the inner planet accretes more than the outer planet for a longer time compared to planets closer to each other. Therefore, the inner planet is starved more easily the closer the planets are.

\begin{figure*}[t]
        \centering   
        \includegraphics[width=18cm]{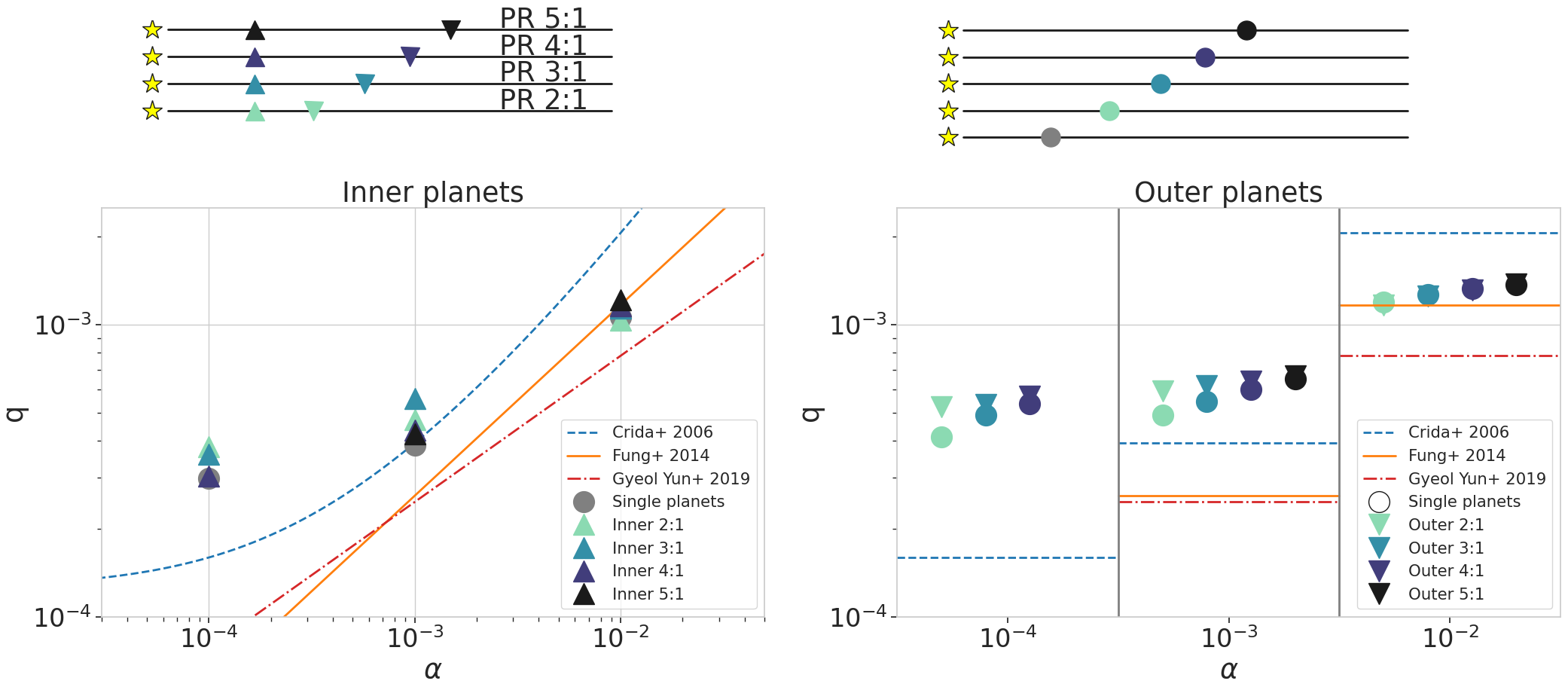}
        \caption[Gap-opening mass as a function of the viscosity for different criteria and our simulations.]{Gap-opening mass as a function of the viscosity for different criteria and our simulations, as in \cite{Bergez2020}. In the left panel, we show the gap opening masses of the inner planets in the two-planet case compared to the single planet cases located at the inner position. The outer planet cases are shown in the right panel. In each panel, the lines represent the different gap-opening criteria from the literature: \cite{Crida2006} in dashed blue, \cite{Fung2014} in solid orange, and \cite{Yun2019} in dashed-dotted red. As shown with the schematics above the panels, the two-planet cases are represented by triangles (upward for the inner planets and downward for the outer planets), and the single planets are represented by circles. The inner single-planet case is shown in gray, and the colored circles correspond to the colors of the outer positions in each configuration. For clarity, in the right panel, the viscosity is divided into discrete intervals, allowing us to compare the gap opening mass at one given viscosity in the different configurations. The gap opening masses of the inner planet are barely impacted by the presence of a second outer planet. On the other hand, at low viscosity, the closer the planets are to each other, the higher the gap opening mass is compared to the single-planet case.}
        \label{2p:2p1p_gap_opening}
\end{figure*}

At low viscosity ($\alpha=10^{-4}$, left panel of Fig. \ref{2p:allconf_mratio_allalphas}), the planet separation has an additional impact on the formation of vortices. In the 2:1 period ratio simulation, the planets quickly create a common gap, preventing the formation of a strong vortex in between them. As we mentioned in Sect. \ref{sec:2p_diffalphas}, as vortices push material toward the planets, it means that the gas accretion of the planets in the 2:1 period ratio are less impacted by the presence of vortices compared to planets located farther away from each other. In the 5:1 period ratio case, the planet gaps are clearly distinct from each other. Each planet therefore create sharp density gradients at the inner and outer edge of their gaps, resulting in the formation of four vortices. Even though the vortices located in the inner disk (i.e., inner to the outer planet) dissipate quickly, they influence the evolution of the planet masses. Except from the formation of vortices, the global behavior of the mass ratio is due to the same processes as at intermediate viscosity.

Overall, independent of the planet separation or disk viscosity, the mass ratios stay close to one (0.7 < $m_{out}/m_{in}$ < 1.25). We discuss in Sect. \ref{sec:2p_discuss_exoplanets} how does this compare to the observed planetary systems.

\subsection{Impact on the gap opening mass}

The gap opening mass is an important parameter used both in theoretical models to approximate when a planet switches from the fast type I migration to the slow type II migration \citep[e.g.,][]{Ndugu2018,Bitsch2019,Miguel2020,Ndugu2021} and in dust observations to indirectly derive the masses of embedded planets from the observed characteristics of gaps \citep[e.g.,][]{Zhang2018,AsensioTorres2021}. As we showed in \cite{Bergez2020}, gas accretion has a non-negligible impact on the gap opening mass. In this section we investigate the influence of the presence of a second accreting planet on the gap opening mass of each planet.

%From \cite{Bergez2020}, we know that the gap opening process depends on the accretion on the planet as well as on the viscosity of the disk: at $\alpha = 10^{-2}$, no large difference is expected as gas accretion has negligible impact on the gap opening mass. At lower viscosities, we expect the gap opening masses to be more similar to the single planet case when the planets are further away from each other. For the outer planets (right panel), at lower viscosities, the difference between the two planets and corresponding single planet cases is smaller when the planets are further away from each other. However, the influence on the inner planets is really negligible: this can be explained by the small difference in accretion between the single and two planet cases as seen on Fig. \ref{2p:2p1p_mdot_mass}.Therefore the presence of a second planet does not highly impact gap opening mass.

We compare in Fig. \ref{2p:2p1p_gap_opening} the gap opening mass of our accreting planets to different gap opening criteria derived in previous studies \citep{Crida2006,Fung2014,Yun2019}. The gap opening masses of the inner planets are shown on the left panel of the figure while the outer planets are shown on the right panel\footnote{The simulation did not reach the gap opening mass in the case of the outer planet of the 5:1 period ratio at low viscosity. The presence of strong vortices prevented the correct continuation of the simulation. However, we clearly see the expected trend.}. As in the previous section, the color represents the different period ratios, as can be seen on the schematics of the disk configurations represented in the top of the figure. Triangles represent the gap opening masses of the planets located in a two-planet disk and single-planet gap opening masses are shown with circles. The gray circles correspond to the gap opening masses of the single planets located at the inner location (they correspond to the gap opening masses presented in \cite{Bergez2020} for the fiducial gas accretion rate). The different lines represent the different gap opening criteria compared in this study. In the right panel, the viscosity is divided into discrete intervals for each of the studied viscosity ($\alpha=10^{-4},10^{-3},10^{-2}$) to help in visualizing the different configurations.

In \cite{Bergez2020}, we conclude that gas accretion has a different impact on the gap opening mass depending on the disk viscosity: at high viscosity, gas accretion helps carve deeper gaps, resulting in a lower gap opening mass for an accreting planet while at low viscosity, gap formation is not helped by gas accretion, resulting in a higher gap opening mass for an accreting planet. When a second planet is added in the disk, the accretion of each planet is impacted by the gap opening of the neighboring planet, as shown in Sect. \ref{sec:2p_2p1pcomp}. However, this impact depends on the viscosity.

At high viscosity, gas accretion helps carve a deeper gap. Here, with $\alpha=10^{-2}$ and $h=0.05$, the disk is at the intersection between the high viscosity regime and the low viscosity regime described above, meaning that the gas accretion has no important impact on the gap opening mass. The gap opening masses are therefore solely dependent on the amount of gas diffusing through the gaps. As the gas diffuses efficiently in this case, gap formation is not impacted by the presence of a second planet. Indeed, the inner disk is more depleted by viscous accretion than by the accretion of both planets, meaning that the material pushed away by the gap forming planets is dissipated via viscous spreading. This can be seen in Fig. \ref{2p:2p1p_gap_opening}, where the gap opening masses are the same in the single or two-planet case, for both the inner and outer planets.

At lower viscosities, gas accretion does not help gap formation \citep{Bergez2020}. Therefore, the gap opening mass depends on the accretion rate of the planet. As presented in Sect. \ref{sec:2p_2p1p_mdotmass}, the accretion rates of the two accreting planets are slightly enhanced compared to the single planets due to the formation of the planetary gaps pushing material in the feeding zones of the neighboring planet. This slight enhancement of the accretion rate of the planets leads to slightly higher gap opening masses. This impacts both the inner and the outer planets, as shown in Fig. \ref{2p:2p1p_gap_opening}. As the gap opening mass then relies on the amount of material pushed toward the neighboring planet, planets that are close enough from each other enhance their gap opening mass more. Intuitively, when the planets are farther from each other, they tend to behave as if they are isolated and have gap opening masses closer to the single-planet simulations.

Overall, the gap opening masses are barely impacted by the presence of a second planet in the disk. The small differences originate from the differences in accretion rates as the planets push material toward each other. Considering the conclusions of \cite{Bergez2020}, it seems that the gas accretion rate on the planet itself has a stronger impact on the gap opening mass than the presence of a simultaneously accreting companion.

\section{Influence of delayed accretion}
\label{sec:delayedaccretion}

Giant planet formation models have very few constraints on the timing at which runaway gas accretion occurs \citep[e.g.,][]{Paardekooper2018,Raymond2020} . So far, we only considered the simultaneous accretion of both giants. However, depending on the disk local properties and on the formation mechanism, giant planets located in the same disk could start accreting at different times. We investigate here the influence of the delayed accretion on the evolution of the planetary growth. Different time delays are considered, on the outer and on the inner planet. We base our delays on the mass of the neighboring planet: the accretion on the second planet is allowed when the other planet reaches 0.3 $M_J$, 0.5 $M_J$ and 1 $M_J$. We chose to investigate the impact of the accretion delay on the 2:1 period ratio configuration at high and intermediate viscosities as they reach these masses in a reasonable computational time. 

\begin{figure*}[t]
    \includegraphics[width=18cm]{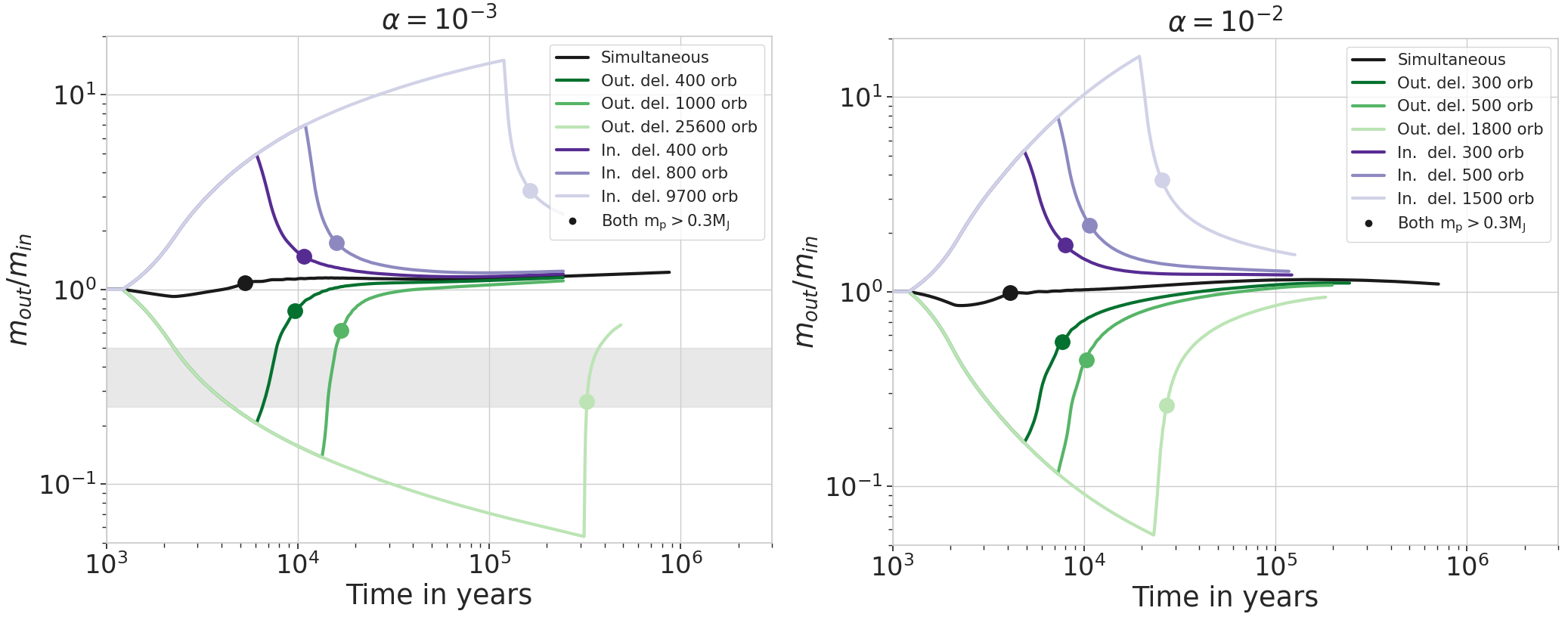}
    \caption[Mass ratio as a function of time for different delays of accretion on the inner or the outer planet.]{Mass ratio as a function of time for different accretion delays on the inner (purple lines) or the outer planet (green lines). The planets are located in the 2:1 period ratio configuration. The two panels represent two different viscosities: $\alpha=10^{-3}$ in the left panel and $\alpha=10^{-2}$ in the right. Darker lines represent shorter delays. Note that this time the mass ratio is displayed on a logarithmic scale (previous plots are on a linear scale) due to the extreme mass ratios induced by our initial setup here. The dots mark the moment when both planets have reached at least 0.3 $M_J$ and can be considered gas giant planets. The gray rectangle represents the region where the mass ratio lies between 0.25 and 0.5 and shows the region where Jupiter and Saturn meet the required conditions to enter common outward migration, for $\alpha \lesssim 10^{-3}$ \citep{Masset&Snellgrove2001,Pierens2014}. Independent of the viscosity, all the mass ratios quickly tend toward the $m_{out}/m_{in}$ = 1 line, leading to similar planet masses in both cases ($0.8 < m_{out}/m_{in} < 2$ after $10^5$ yrs). }
    \label{2p:delayed_mratio}
\end{figure*}

All the resulting mass ratios are shown in Fig. \ref{2p:delayed_mratio}. The mass ratio of planets simultaneously accreting is shown and corresponds to the 2:1 mass ratio presented in Fig. \ref{2p:allconf_mratio_allalphas}. We note the difference in the mass ratio scale: while before the mass ratios are shown on a linear scale, here the scale is logarithmic for readability. The large spread in mass ratio is induced by our initial choice for the different delays. Indeed, as we wait for the neighboring planet to reach a given mass before accreting, this sets the maximal and minimal mass ratio reached by the planets: the maximal value that is reached is 16 ($318/20 M_\oplus$) and the minimal one is 0.06 ($20/318 M_\oplus$). We therefore expect the planets to reach a final mass ratio located in between these initial values. 

The evolution of the different mass ratios is very different from the simultaneously accreting planets. When the outer planet accretion is delayed (green lines in Fig. \ref{2p:delayed_mratio}), it allows the inner planet to accrete slightly more gas before being starved by the growth of the outer planet. At high viscosity, the gas diffuses efficiently through the planet gaps, depleting the whole disk in gas and leading to a high stellar accretion rate (see Sect. \ref{sec:2p_stellaracc}). Therefore, a longer delay results in accretion in a more depleted disk. As a consequence, the mass ratio is lower for longer delays. 

At lower viscosity, the effect of viscous spreading as described above can be perturbed by the formation of the inner planet gap. However, the gap opening mass at this kinematic viscosity is around 0.5 $M_J$ at the location of the inner planet. For a delay of 300 orb. ($\sim 4.8\times10^3$ yrs), corresponding to an inner planet mass of 0.3 $M_J$, the outer planets starts accreting while the inner planet did not create a deep gap yet. This leads to a very similar final mass ratio evolution as in the simultaneous case. However, we know from Sect. \ref{sec:2p_massratiosdiffnu} that when the inner planet creates its gap, it starts to deplete the material located in the inner disk and in between the planets. Therefore, when the outer planet starts accreting even later (e.g., when the inner planet has reached its gap opening mass), the inner planet already depleted part of the material in between them, leading to a lower gas accretion rate of the outer planet compared to the simultaneous case. This results in lower mass ratios for longer delays.

The mass ratio of the planets in the case of the delayed accretion of the inner planet (purple lines of Fig. \ref{2p:delayed_mratio}) highly depends on the depth of the gap of the outer planet and on the viscosity of the disk. Indeed, at low viscosity, if the inner planet starts accreting before the formation of the gap of the outer planet, then the inner region is not depleted in gas yet. The behavior of the mass ratio then quickly tends to be the same as in the simultaneous case. However, if the outer planet already opened its gap, then a longer delay of accretion results in a more depleted inner disk. The inner planet has therefore less material to accrete, leading to higher mass ratios. At high viscosity, the same behavior occurs, with the inner disk being efficiently depleted by stellar accretion.

Even if different mechanisms influence the mass ratio in the case of delayed accretion, the giant planets always end up with rather similar masses: $0.8 < m_{out}/m_{in} < 2$ after $10^5$ yrs. We discuss in Sect. \ref{sec:2p_discuss_exoplanets} how does this compare to the observed mass distributions in the exoplanet population and what kind of constrains on planet formation can be derived.

\section{Discussion}
\label{sec:2p_discuss}

\subsection{Accretion onto the star}
\label{sec:2p_discuss_ontostar}

In Sect. \ref{sec:2p_stellaracc} we investigate the influence of the presence of multiple gas accreting planets on the stellar accretion at different viscosities. The results are compared to the stellar accretion in disks hosting single planets and in disks without any planet. We find that the presence of the second planet only has a significant effect when the viscosity of the disk is low ($\alpha \lesssim 10^{-4}$). For higher viscosities, the presence of the second planet only influences the stellar accretion rate by up to $30\%$ compared to the case with single planets. 

We compare our results with two planets to the different stellar accretion rates obtained with different planetary accretion rate in \cite{Bergez2020}. At high viscosity, the presence of the second planet has less impact on the stellar accretion rate compared to a significant enhancement of the planet accretion rates. Due to the uncertainties in gas accretion rates \citep[e.g., spread in the values found in the following studies:][]{Kley1999,DAngelo2003,Tanigawa2007,Ayliffe2009,Machida2010,Tanigawa2016,Crida2017,Schulik2019,Lambrechts2019} , it is impossible to use the stellar accretion rates to determine if the protoplanetary disk hosts a single fast accreting planet or multiple planets accreting at a lower rate. 

As discussed in \cite{Bergez2020}, our results are quite different from the study derived by \cite{Manara2019}. In their models, they find that the stellar gas accretion rates can be reduced by over two orders of magnitude when the disk is hosting accreting giant planets. We showed in our previous paper that these large spreads of stellar accretion rates could only be reached by widely changing the disk viscosity (over several orders of magnitude). With this study, we additionally show that the presence of a second accreting companion cannot explain such a large reduction in the stellar accretion rate either. Indeed, while reducing the disk viscosity to $\alpha = 10^{-4}$ enhanced the impact of the two accreting planets on the stellar accretion rate, we expect the reduction to be of a factor of 10 at most compared to a disk without planets. These discrepancies between our study and the work done by \cite{Manara2019} are the same as mentioned in our previous paper: their model simulates a 1D gas disk while our study is performed in 2D, allowing us to more accurately determine the flow of gas through the gaps of the planets \citep[e.g.,][]{Lubow2006}. Moreover, their planetary gas accretion rates might be overestimated as they rely on the unperturbed surface density \citep{Mordasini2012}, while we showed here that the depletion of the inner disk leads to the starvation of the inner planet and consequently a reduction of its accretion rate.

Again, our simulations indicate that planetary gas accretion might have a smaller impact than expected on the stellar gas accretion rates, even in the presence of multiple accreting planets.

\subsection{Dependence on the accretion behavior}

The results from our simulations rely on our accretion recipe, based on the results from \cite{Machida2010} and on the amount of gas available around the planets (i.e., the total mass of the disk and the gas surface density profile). However, one can argue that our understanding of planetary gas accretion in the runaway phase can be improved. In the current literature, the derived accretion rates can range over several orders of magnitude, ranging from around $10^{-8}$ $\rm M_J/yr$ \citep{Tanigawa2007,Tanigawa2016} up to around $10^{-4}$ $\rm M_J/yr$ \citep{DAngelo2003,Schulik2019}. Furthermore, some studies \citep[e.g.,][]{Lambrechts2019} show that the actual accretion rate has to be separated from the flux that goes through the atmosphere as the planet is embedded in the gas disk. With this definition, the flux through the atmosphere is on the order of $10^{-4}$ $\rm M_J/yr$ but the actual accretion rate is lower, around $10^{-6}$ to $10^{-5}$ $\rm M_J/yr$. 

However, in our study, the global evolution of the planet's mass ratio relies mostly on the evolution of the global gas distribution in the disk, via gap opening and the depletion of the disk. Therefore, the way the disk is evolving has a larger impact than the actual accretion rate.

As mentioned in Sec. \ref{sec:2pnumericalsetup}, we start our simulations with a relatively massive disk with a surface density profile following $\Sigma \propto r^{-1}$. Implanting the planets early implies that a large amount of gas is still available for the planets to accrete. If we wait for a longer time before introducing the planetary cores, the disk has time to be accreted onto the star and therefore less material would be available around the planets. This will lead to a lower planetary accretion rate on one hand, and on less material between them on the other hand. Then it is just a question of timing again between the formation of the gaps and the depletion of the inner disk and the material between the planets: the global evolution of the planet mass ratio is expected to be the same but on longer timescales as the gap opening mass is independent of the disk's surface density \citep{Crida2006,Fung2014,Kanagawa2015}. Therefore, the timing of formation (and therefore the total mass of the disk) will be of importance regarding the total mass reached by the planets but will not have a huge impact on the evolution of their mass ratio. Regarding the importance of the surface density profile, the evolution of the planets' mass ratio will also be shifted in time but the qualitative evolution is expected to be similar.

\subsection{Impact of planet dynamics}
\label{sec:2p_planetdynamics}

In order to determine the impact of the gas accretion on two planets embedded in the same disk, we neglected both the dynamical interactions between the planets and their migration. Regarding migration, different studies investigate how it impacts gas accretion \citep[e.g.,][]{Durmann2015,Crida2017,Durmann2017}. The main results of these studies are that the evolution of the planet characteristics (i.e., mass and semimajor axis) highly depends on the timescales of each process: a fast migrating planet tends to accrete more gas as it quickly moves toward regions with high surface densities. However, as we showed in \cite{Bergez2020}, gas accretion has an impact also on the gap opening mass, influencing the migration speed of the planet as it transitions from a fast type I to a slow type II migration. This effect was observed in \cite{Crida2017}, where their accreting planet slowed down its migration speed earlier compared to a non accreting planet. Therefore, the resulting planetary systems highly depend on the timescales of migration, gap formation and gas accretion. 

Another dynamical effect can play an important role in the evolution of the planets. Previous hydrodynamical studies show that multiple planets can be captured in resonant chains during the gas phase of the disk \citep[e.g.,][]{Baruteau2013,Pierens2014,Kanagawa2020}. The capture in resonance can have an important impact on the migration behavior of the planets. For example, in the Grand Tack scenario \citep[e.g.,][]{Masset&Snellgrove2001,Walsh2011,Pierens2014}, Jupiter and Saturn are believed to migrate inward and then outward due do their capture in resonance. This outward migration is occurring for precise disk parameters and mass ratios, as shown in \cite{Pierens2014}. As migration can be altered by the capture in resonance, gas accretion and gap formation will also be indirectly altered by the planet radial motion.

The capture in resonance can lead to a slight increase of the planets eccentricity. As an eccentric planet will open a less deep gap for the same mass \citep[e.g.,][]{Hosseinbor2007,SanchezSalcedo2022}, the planet will end up with higher mass accretion rates than our planets on circular orbits. Having a higher accretion rate will just result in a shift in the different timing but this impact is negligible for low eccentricities. Since the eccentricity of the planets is damped while the amount of gas is significant in the disk \citep[e.g.,][]{Moorhead2009,Bitsch2010}, and since our planets do not grow massive enough to excite their eccentricity \citep[more than 5 $M_J$; e.g.,][]{Papaloizou2001,Kley2006,Bitsch2010,Bitsch2013eccentricity}, we expect the impact of realistic low eccentricities on our results to  not be significant.

We plan on implementing the impact of both migration and capture in resonance on the growth of our two planets in follow-up studies. We expect that, if the planets start accreting simultaneously, then the structure of the resulting system highly depends on the timing at which gap opening will occur because it will slow down the migration of the planets and determine their accretion behavior. A potential outward migration can delay the depletion of the inner disk, altering the mass ratio behavior discussed in Sect. \ref{sec:2p_massratiosdiffnu}. However, due to the high interdependence of each mechanism, it is difficult to precisely predict how the planets will behave.

\subsection{Implications for the Grand Tack scenario}
\label{sec:2p_grandtack}

In the Grand Tack scenario, if Jupiter and Saturn have a mass ratio between 0.25 and 0.5, then they can migrate outward to their current locations \citep[e.g.,][]{Masset&Snellgrove2001,Crida2009,Pierens2014}. In Sect. \ref{sec:2p_mratio_allconf}, we show that if the planets start accreting simultaneously, they reach mass ratios that are between 0.7 and 1.3. Therefore, in order to trigger outward migration, the planets have to start accreting with a delay. It also requires that the inner planet is more massive than the outer planet; otherwise, the torques arising from the outer disk would be too large, leading to inward migration. This also implies that Jupiter must have started to accrete gas efficiently before Saturn.

\cite{Pierens2014} find that outward migration depends also on both the period ratio of the planets and the disk parameters. In order to trigger outward migration in a low mass disk, a capture in a 2:1 resonance is needed. If the disk is more massive, then the planets need to reach the 3:2 resonance. Both scenarios require a relatively low $\alpha$-viscosity ($\alpha \lesssim 10^{-3}$). 

Within our current parameter study, we only investigated the impact of delayed accretion in the 2:1 period ratio configuration. At low viscosity ($\alpha=10^{-3}$,left panel of Fig. \ref{2p:delayed_mratio}), the conditions are barely met for the outward migration to occur: in all cases, the mass ratios quickly evolve in the disk, making the planets barely stay in the needed mass ratio range (marked by the gray area). Therefore, with our current results, it seems that outward migration of the two giant planets is very challenging to reach as this occurs during a very short timescale. However, we plan to expand our parameter space study in the near future, allowing us to better analyze if and how a planetary system like Jupiter and Saturn could have formed via the Grand Tack scenario.

\subsection{Comparison to exoplanets}
\label{sec:2p_discuss_exoplanets}

\begin{figure}[t]
        \centering   
        \includegraphics[width=9cm]{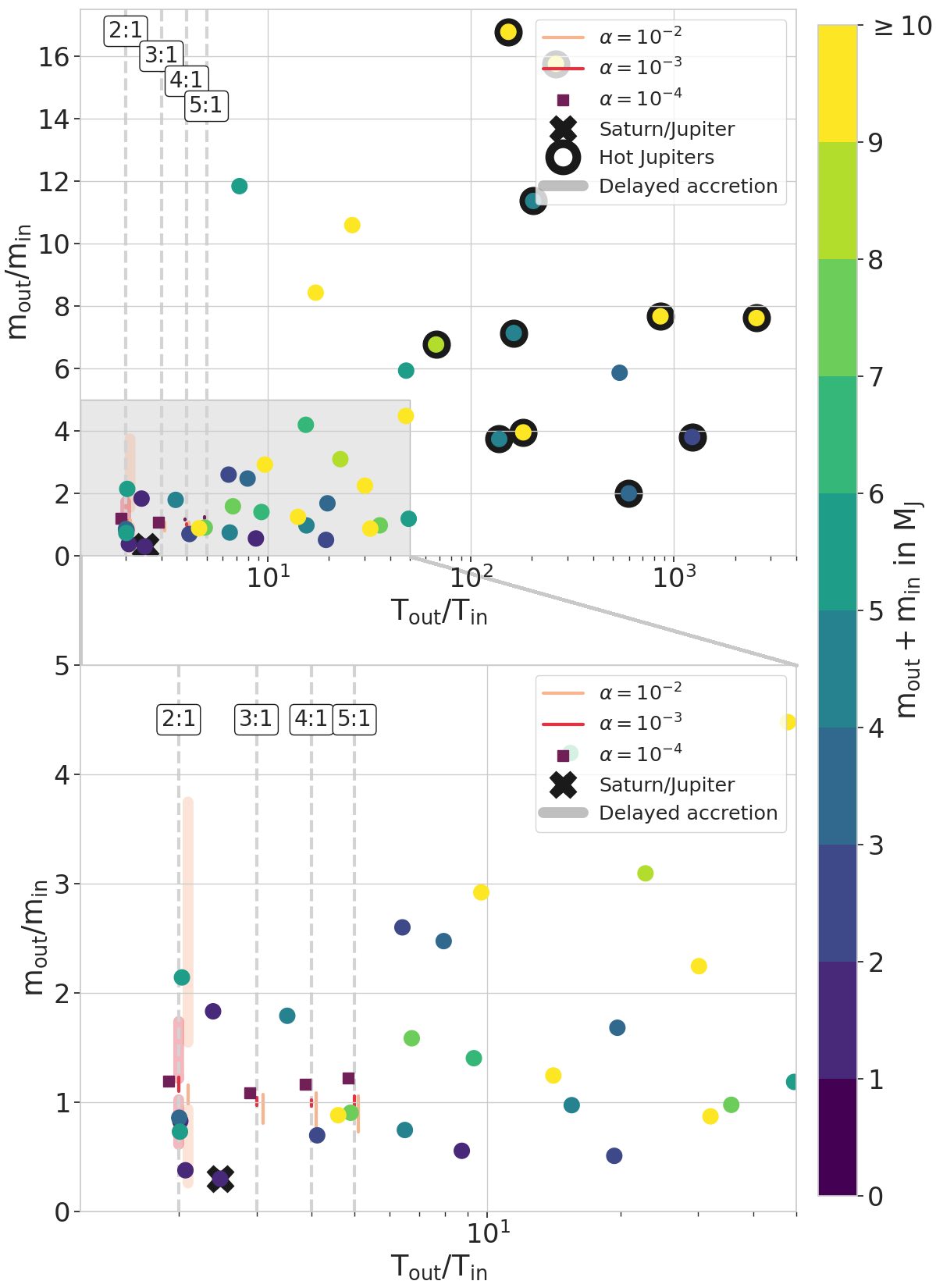}
        \caption[Mass ratio evolution of the two accreting planets as a function of their period ratio compared to exoplanetary systems.]{Mass ratio evolution of the two accreting planets as a function of their period ratio compared to exoplanetary systems. The investigated period ratios from our simulations are marked by vertical dashed gray lines. The data are taken from the NASA exoplanet archive, for which we selected the systems as follows: the system contains exactly two detected planets, both of them larger than 0.3 $M_J$. They orbit a single Sun-like star ($4700K<\mathrm{T_{eff}}<6500K$ and $\log(g_*) > 4$). The color of the dots represents the sum of the planet masses in Jupiter masses, without error bars. A black contour surrounds planets that are considered hot Jupiters (i.e., with a period of less than 10 days). As we expect their formation to be highly influenced by the dynamics of the system, which we do not model here, we focus the comparison with the exoplanets on the planets marked by the gray area in the top part of the figure. A zoomed-in view of this region is shown in the second panel. The vertical red and pink lines represent the maximum and minimum mass ratio reached in each of our simulations once both planets reach 0.3 $M_J$. Darker colors represent lower viscosities. For visibility, the lines corresponding to the different $\alpha$-viscosities are slightly offset from the period ratio line. The extent of the $\alpha = 10^{-4}$ was so small at low period ratios that we represent it with squares. The cross marks the Jupiter and Saturn couple. Simultaneously accreting planets lead to planets that are very similar in mass compared to the exoplanet population. Some systems seem to be consistent with simultaneous accretion; however, another mechanism is needed to explain the existence of the other systems. Delayed accretion as shown in Fig. \ref{2p:delayed_mratio} coupled with different disk lifetimes could explain the difference in planetary masses.}
        \label{2p:comp_exo}
\end{figure}

Considering our current parameter space study, planets accreting from the same disk end up with very similar planet masses. Delaying the accretion of the respective planets allowed us to slightly broaden the mass ratio range reached; however, in $10^5$ yrs, the final mass ratios obtained are still between 0.8 and 2. These mass ratios are quite different from the mass ratios observed in different planetary systems. In Fig. \ref{2p:comp_exo}, we compare the evolution of our mass ratios to different exoplanetary systems. The data originate from the NASA exoplanet archive\footnote{https://exoplanetarchive.ipac.caltech.edu/docs/data.html}. We selected the planetary systems as follows: first of all, we are interested in systems containing exactly two giant planets (i.e., with $m_p > 0.3 \; M_J$) as we investigate the accretion of two planets in the runaway gas accretion phase. Our simulation considers planet formation in a disk orbiting the Sun; therefore, the selected planetary systems orbit Sun-like stars ($4700K<\mathrm{T_{eff}}<6500K$ and $\log(g_*) > 4$). Each panel represents the ratio of the outer planet mass divided by the mass of the inner planet like in previous figures as a function of the planet period ratio. Vertical dashed lines represent the investigated period ratios. The colorbar shows the sum of the planet masses in $M_J$.

From the top panel in the figure, it is clear that the observed planetary systems hosting two giant planets have a broad range of mass and period ratios. We highlight the systems hosting at least one hot Jupiter (i.e., planets with periods shorter than 10 days) with a thick black contour. These planets might be formed after a very efficient inward migration or via a scattering event, leaving them very close to their host star. This results in a system where the period ratio of the planet is very large. As we do not implement migration in this study, we focus on the planets located closer to each other. Interestingly, it appears that planets located closer to each other seem to have more similar masses (except for 3 systems with mass ratios higher than 8). For better readability and comparison with our simulation, a zoom on the planets placed in the gray rectangle is shown in the lower panel of the figure. 

Due to computational constraints, we could not investigate the evolution of the mass ratio until the end of the disk lifetime. Therefore, in order to make the comparison with fully formed planets as the ones observed in the different planet surveys, we show the maximal and minimal mass ratios obtained by our simultaneously accreting planets with the pink, red and purple vertical lines. As we selected the observed systems by considering planets with $m_p > 0.3\;M_J$, we show the mass ratio spread once both planets reached 0.3 $M_J$. The ratios obtained for different viscosities are slightly offset from the exact period ratio for visibility. At low viscosity and small period ratios, the mass ratio range was too small to be represented by a line; therefore, the final mass ratio is shown with a square. Over the 45 observed planetary systems (including Jupiter and Saturn and hot Jupiters), 5 have a mass ratio lower than 0.7, 11 systems have a mass ratio lying between 0.7 and 1.3 and 29 systems have a mass ratio higher than 1.3. The majority of the systems cannot be explained by the simultaneous accretion of the planets. We also note that very few systems (only 5 here) feature a mass ratio as seen in our Solar System with Jupiter and Saturn. While this might be explained by the difficulty of our current facilities to see low mass planets, it also raises the question of the peculiarity of the Solar System among other systems, namely whether our planetary system common or an outlier.

 In this study, such a large spread in mass ratio was only reached when the planets accrete in the runaway gas accretion phase with different delays. This accretion delay can be justified by the dependence of the beginning of the runaway gas accretion phase on the disk characteristics. Here, we followed the classical core accretion model, where it is assumed that runaway gas accretion is triggered when the planetary core reaches a mass of 20 $M_\oplus$, with a solid core of 10 $M_\oplus$ surrounded by a first gaseous atmosphere of 10 $M_\oplus$ \citep{Pollack1996}. However, more recent studies show that the initial total mass of the core can vary depending on the local properties of the disk. Runaway gas accretion can be triggered at different core masses depending on the local disk temperature and opacity \citep{Ikoma2001,Piso2014,Bitsch2021}. Moreover, in the pebble accretion scenario, the pebble isolation mass corresponds to the mass at which the core is shielded from the pebble flux by the pressure bump created by its own gap \citep{Morbidelli2012,Lambrechts2014,Bitsch2018a,Ataiee2018}. Then, the atmosphere of the planetary core is not heated anymore by the accretion of solids and cools down, entering the runaway gas accretion phase. The pebble isolation mass depends on the disk's aspect ratio, $\alpha$-viscosity, pressure profile and turbulent diffusion of the particles \citep{Bitsch2018a}. Therefore, the delay of accretion time between two giants highly depends on the local properties of the disk. Protoplanetary disks can be flared, featuring large aspect ratio variations \citep[e.g.,][]{Bitsch2015a,Pierens2016}, and different hydrodynamical properties can lead to important radial variations of viscosity \citep[e.g.,][]{Flock2011,Dullemond2018b,Delage2022}. The disk properties can lead to the delay of either the inner or the outer planet. 

Even when we applied different accretion delays, the long-term trends could only reproduce mass ratios lying between 0.8 and 2. With these parameters, the only remaining way to reach large (or small) mass ratios, is to stop the accretion at a given mass ratio. The timing of the dissipation of the gas disk can be crucial here. For example, photoevaporation can dissipate the gas disk from inside out by creating a inner hole separating the inner disk ($r<1$ AU) from the outer disk and quickly depleting it \citep[for a review, see][]{Ercolano2017}. Such a depletion of the disk might have the capacity to starve the giant planets, influencing the evolution of their mass ratio.

Some differences between our simulations and the observations might originate from dynamical events that are not yet included in our study. For example, dynamical interactions such as collisions might change the planets mass ratio \citep[e.g.,][]{Juric2008,Raymond2009a,Sotiriadis2017,Bitsch2020}. We also note that planets might also be ejected from the systems after the dispersal of the gaseous disk: while the observed final system would host two giants, the gaseous protoplanetary disk where they formed could have hosted three giants, changing the accretion behavior compared to what we simulate in this paper. However, accurately describing the hydrodynamical evolution of the gas together with the dynamical evolution of more than two planets for a long time (over several Myrs) is quasi-impossible nowadays. Therefore, further studies should slowly including more dynamical effects during the gas phase or slowly improving the hydrodynamical assumptions of N-Body simulations.

To summarize, the simultaneous runaway gas accretion of fixed planets cannot explain the distribution of exoplanetary masses observed as it leads to planets with very similar masses. To increase the difference in planet masses, the accretion between the planets have to be delayed and efficient disk dispersal mechanisms are required to end the growth of the planets at given mass ratios. These two last points highly depends on the disk local properties of the gas. Moreover, taking dynamical interactions between the planets themselves and the planets and the disk might improve our understanding of the giants distribution. A future study including dynamical interactions and migration is planned to determine how the results can be impacted.

\section{Conclusions}
\label{sec:2p_conclusion}

In this paper we investigate the mass distribution of two accreting planets located in the same disk. Using 2D hydrodynamical simulations, we monitor the evolution of the planetary mass ratio for different disk viscosities, different planet configurations, and different accretion timings. Our main conclusions can be summarized as:

\begin{enumerate}
    \item The evolution of multiple accreting planets is mainly governed by the viscosity of the disk. The mass ratio evolution of simultaneously accreting planets depends on the balance between the gas accretion and gap opening timescales. As shown in \cite{Bergez2020}, at high viscosity, when gas accretion acts in favor of gap formation, the inner planet is rapidly starved by the viscous accretion onto the star and the outer planet accretes more until becoming more massive than the inner planet. 
    
    However, at lower viscosities, when gap formation is only dependent on the disk reaction time (i.e., when gas accretion does not help gap formation), the evolution of the mass ratio of the planets follows a different behavior: the outer planet accretes more gas until the inner planet forms its gap. Then, the inner planet starts depleting the inner disk and the material present in between the planets, resulting in a higher accretion rate for the inner planet than for the outer one. When the amount of material located in these two regions is significantly depleted, the inner planet becomes starved by the outer planet. 
    
    \item Simultaneously accreting planets always end up with similar masses, independent of the disk viscosity. In order to reach more extreme mass ratios, we simulated a delayed accretion of the inner or outer planet in one configuration at high and intermediate viscosities. While the initial mass ratios are large by construction, the planets quickly tend toward similar mass ratios ($0.8 < m_{out}/m_{in} < 2$ in $10^5$ yrs). 
    
    \item Via comparisons with the observed exoplanet population, we conclude that gas accretion occurring at the same time can explain the characteristics of only a few planetary systems. Delayed accretion coupled with different disk lifetimes leads to mass ratios that are more consistent with the observations. While core formation timescales and different disk dissipation mechanisms can explain the possibility of a delayed accretion and depletion of gas, our study shows that the majority of the observed systems of multiple gas giant planets should have started to accrete at different times.
    
\end{enumerate}

Understanding how material is distributed between multiple planets is crucial to better understanding the dynamical evolution of the forming system. As discussed in Sect. \ref{sec:2p_planetdynamics}, the radial evolution of multiple planets is governed by the migration and capture in resonance of the planets, themselves dependent on the gas distribution in the disk, which in turn is governed by gap formation and by planetary and stellar gas accretion. Future studies investigating the growth of multiple giant planets should both consider that the gas accretion of the planets is impacted by the presence of neighboring planets and provide mechanisms that explain the spread in mass ratios observed in distant exoplanetary systems as well as our own Solar System.

\begin{acknowledgements}
      C. Bergez-Casalou and B. Bitsch thank the European Research Council (ERC Starting Grant 757448-PAMDORA) for their financial support. S.N. Raymond thanks the CNRS's PNP program.
\end{acknowledgements}

\bibliographystyle{aa}
\bibliography{biblio}

\begin{appendix}
\section{Gas accretion routine}
\label{appendix:gas_accretion_routine}

As in \cite{Crida2017} and \cite{Bergez2020}, gas accretion is modeled using the \cite{Machida2010} and \cite{Kley1999} principles. The \cite{Machida2010} accretion rate is derived by fitting 3D isothermal shearing box simulations and corresponds to the runaway gas accretion phase of the giant planet. This accretion rate is different than in the first slow contraction phase, as predicted by the core accretion model \citep{Piso2014}. It can be written as
\begin{equation}
    \dot{M}_{M}=<\Sigma>_{0.9r_H} H^2\Omega_p \times min\Big[0.14;0.83(r_H/H)^{9/2}\Big]
    \label{eq:mach_acc}
,\end{equation}
where $ r_H = r_p(m_p/3M_*)^{1/3}$ is the Hill sphere of the planet, $<\Sigma>_{0.9r_H}$ is the averaged local surface density from the planet location up to $0.9 r_H$; $H$ is the disk scale height and $\Omega_p$ is the Keplerian orbital frequency of the planet. 

On the other hand, in order to make sure that the planetary accretion is limited by what the disk can provide, we consider that the maximal accretion rate of the planet is given by the \cite{Kley1999} principle. This arbitrary accretion rate considers that the planet can accrete a given fraction of the gas present in its Hill sphere. This fraction of gas is dependent on the distance to the planet and is given by $f_{\rm red}$. This accretion rate can be written as

\begin{equation}
      \dot{M}_{K}= \iint_{A_{disk}} f_{\rm red}(d)\;\Sigma(r,\phi,t)\;\pi f_{\rm acc} \; dr \; d\phi
    \label{eq:kley_acc}
,\end{equation}
where $A_{disk}$ is the area of the disk, $f_{\rm red}$ is a smooth reduction function predicting the fraction of gas accreted by the planet as a function of the distance from the planet $d$, $\Sigma(r,\phi,t)$ is the gas surface density around the planet and $f_{\rm acc}$ is the inverse timescale upon which the accretion is occurring. $f_{\rm red}$ is the smooth reduction function used to predict what fraction of gas can be accreted on the planet as a function of the distance to the planet $d$. It is defined as\begin{equation}
    f_{\rm red} =
    \begin{cases}
        2/3 & \text{if} \; d<0.45 \; r_H \\
        2/3 \times \cos^4\Big(\pi\Big(\frac{d}{r_H}-0.45\Big)\Big) & \text{if} \; 0.45\; r_H<d<0.9 \; r_H
    \end{cases}
    \label{eq:fred}
.\end{equation}

This function is based on \cite{Robert2018}, where the authors assume that close to the planet gas accretion is $100\%$ efficient ($\rm f_{\rm red} = 1$). However, a $100\%$ efficiency is not realistic in such a case, as shown by \cite{Schulik2019}. The accreted mass fraction increases closer to the planet but it does not accrete $100\%$ of the gas in the vicinity. Thus, we chose to limit our study to a maximum  $\rm f_{\rm red}$ value of 2/3.

By combining these two methods, the final amount of gas that the planet can accrete is determined by $\dot{M}_p = min(\dot{M}_M,\dot{M}_K)$. The gas is removed from the disk and added to the mass of the planet. To do so, we remove the gas in this regime with the same formalism as for the \cite{Kley1999} method. This means that if the planet is accreting in the regime of \cite{Machida2010}, the total amount of gas it accretes is given by Eq.\ref{eq:mach_acc} and the distribution for where the gas is removed is given by Eq.\ref{eq:fred}.

This way, the removal scheme of the gas is the same for both principles, but the mass that can be accreted is limited either by the derived accretion rates of \cite{Machida2010} or by the maximum amount the disk can provide, given by the \cite{Kley1999} method. In this study, we remain in the regime where $\rm \dot{M}_M < \dot{M}_K$ throughout, meaning that we always remove the amount of mass dictated by \cite{Machida2010}.

\section{Surface density maps}
\label{appendix_2Dgas_2p}

\begin{figure*}[t]
        \centering   
        \includegraphics[width=18cm]{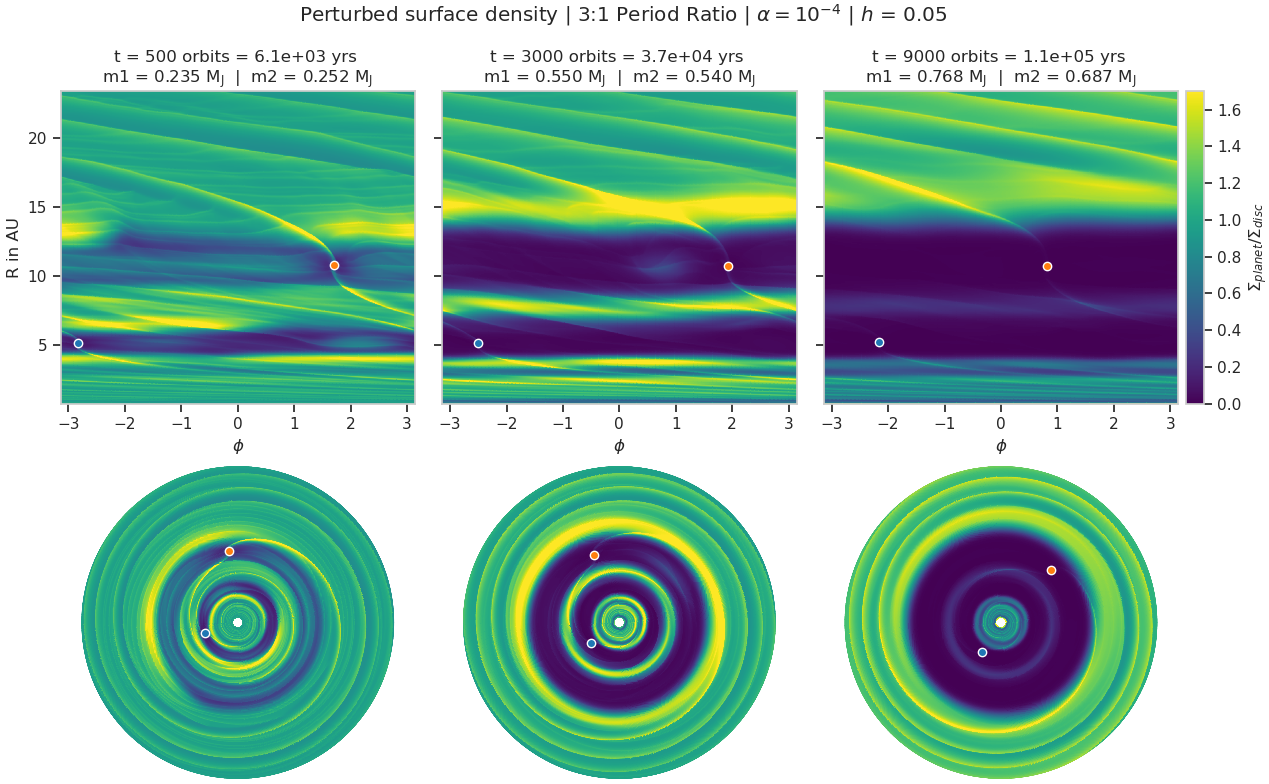}
        \caption[2D perturbed surface density maps at three different times to trace the evolution of vortices]{2D perturbed surface density maps at three different times: t = 500 (left), 3 000 (middle), and 9 000 (right) inner planet orbits. The over-densities (yellow asymmetries) are representative of vortices.}
        \label{2p:vortices}
\end{figure*}

In Sect. \ref{sec:2p_diffalphas}, we investigate the influence of the disk viscosity on the evolution of the planets growth. At low viscosity ($\alpha = 10^{-4}$ and $h=0.05$), the RWI \citep{Lovelace1999,Li2001} is triggered at the edges of the different planet gaps, creating vortices. In Fig. \ref{2p:vortices}, we show the 2D perturbed surface density maps of the disk hosting two accreting planets in the 3:1 period ratio, at three different times, increasing from left to right. Polar plots of the density maps presented in the top row are shown in the bottom row. The vortices produce asymmetric over-densities that can be used to trace them (in yellow in Fig. \ref{2p:vortices}). 

At the beginning of the simulation (at 500 inner planet orbits, left panel), we see that vortices are produced at three different locations in this configuration: at the outer edge of the outer planet gap, in between the planets and interior to the inner planet gap. Their presence impacts the flow of gas in the vicinity of the planets, creating oscillations in the planetary accretion rates (see Sect. \ref{sec:2p_diffalphas}). Quickly, the vortices located in between the planet and in the inner disk vanish (middle panel). The strongest vortex (i.e., with the largest over-density) is the one located at the outer edge of the outer planet gap. It takes longer to dissipate and is completely vanished after $10^5$ yrs (right panel). The strength of the vortices depends on the growth timescale of the planets \citep{Hammer2017}: if the planets accrete faster, the vortices would be stronger but would also vanish faster. 

\end{appendix}

%-------------------------------------------------------------------

\end{document}